\documentclass[%
superscriptaddress,
amsmath,amssymb,aps,prd,twocolumn
]{revtex4-2}

\usepackage{graphicx}
\usepackage{dcolumn}
\usepackage{bm}
\usepackage{xspace}
\usepackage{xcolor}
\usepackage{mathtools} 
\usepackage{hyperref}
\definecolor{darkblue}{rgb}{0.18,0.19,0.57}
\hypersetup{
    colorlinks,
    linkcolor=darkblue,
    citecolor=darkblue,
    urlcolor=darkblue
}

\newcommand{\im}{\ensuremath{\mathrm{i}}\xspace}

\newcommand{\sign}[1]{\ensuremath{\mathrm{sgn}(#1)}\xspace}
\definecolor{darkgreen}{rgb}{0.0, 0.5, 0.0}

\def\vec#1{{\bf{#1}}}
\newcommand{\nn}{\nonumber\\}

\newcommand{\f}[1]{\mbox{\boldmath$#1$}}
\newcommand{\fk}[1]{\mbox{\boldmath$\scriptstyle#1$}}

\newcommand{\bea}{\begin{eqnarray}}
\newcommand{\ea}{\end{eqnarray}}
\newcommand{\eea}{\end{eqnarray}}
\newcommand{\ord}{{\cal O}}

\begin{document}

\title{Ginzburg effect in a dielectric medium with dispersion and dissipation}

\author{Sascha Lang} \email{s.lang@hzdr.de} 
\affiliation{%
 Helmholtz-Zentrum Dresden-Rossendorf, Bautzner Landstra{\ss}e 400, 01328 Dresden, Germany
}
\affiliation{%
 Fakult\"at f\"ur Physik, Universit\"at Duisburg-Essen, Lotharstra{\ss}e 1, 47057 Duisburg, Germany
}

\author{Roland Sauerbrey}
\affiliation{%
 Helmholtz-Zentrum Dresden-Rossendorf, Bautzner Landstra{\ss}e 400, 01328 Dresden, Germany
}
\affiliation{
Institut f\"ur Angewandte Physik, Technische Universit\"at Dresden, 01062 Dresden, Germany}
\affiliation{
Center for Advanced Systems Understanding (CASUS), 02826 Görlitz, Germany}

\author{Ralf Sch\"utzhold}%
\affiliation{%
 Helmholtz-Zentrum Dresden-Rossendorf, Bautzner Landstra{\ss}e 400, 01328 Dresden, Germany
}%
\affiliation{%
 Institut f\"ur Theoretische Physik, Technische Universit\"at Dresden, 01062 Dresden, Germany
}
\affiliation{%
 Fakult\"at f\"ur Physik, Universit\"at Duisburg-Essen, Lotharstra{\ss}e 1, 47057 Duisburg, Germany
}
\author{William G. Unruh}
\affiliation{
 Department of Physics and Astronomy, University of British Columbia, Vancouver V6T 1Z1, Canada
}%
\affiliation{
 Institute for Quantum Science and Engineering, Texas A\&M University, College Station, 
 Texas 77843-4242, USA
}%

\date{\today}

\begin{abstract}
As a quantum analog of Cherenkov radiation, an inertial photon detector moving through 
a medium with constant refractive index $n$ may perceive the electromagnetic quantum 
fluctuations as real photons if its velocity $v$ exceeds the medium speed of light $c/n$.
For dispersive Hopfield type media, we find this Ginzburg effect to extend to much lower 
$v$ because the phase velocity of light is very small near the medium resonance. 
In this regime, however, dissipation effects become important. 
Via an extended Hopfield model, we present a consistent treatment of quantum fluctuations 
in dispersive and dissipative media and derive the Ginzburg effect in such systems.
Finally, we propose an experimental test.
\end{abstract}

\maketitle

\section{Introduction} 

As already realized by S.~Fulling half a century ago, the particle concept in relativistic quantum 
field theories is not unique~\cite{Fulling_1973}. 
This non-uniqueness lies at the heart of many phenomena, such as Hawking 
radiation~\cite{HawkingNature_1974, HawkingComm_1975}, 
cosmological particle 
creation~\cite{Schroedinger_1939, Parker_1968, Parker_Interview} 
and acceleration radiation 
(a.k.a.~the Unruh effect~\cite{Unruh_1976, Unruh_1984}).
Adopting a quite pragmatic point of view, one could say that a particle is what makes a particle 
detector click.
Even though this appears as a tautology, it summarizes the idea that particles 
are defined via their interplay with matter (e.g., in the form of detectors).

An inertial detector perceives the Minkowski vacuum as empty (of particles) 
because energy conservation implies that the detector must absorb a quantum 
from the field (which does not exist in the vacuum) in order to click. 
In more formal terms, the quantum vacuum fluctuations only contain positive frequencies 
with respect to the detector's proper time (consistent with the \emph{spectrum condition}
of the Wightman function~\cite{Wightman_1956, Strocchi_2004, WightmanStreater_1964}).

However, a non-zero response may occur, for instance, for a uniformly accelerated detector,
which perceives the Minkowski vacuum as a thermal bath~\cite{Unruh_1976}
because the quantum vacuum fluctuations, translated to the detector's 
proper time, also contain negative frequencies 
\footnote{
Note that, quite generally, the creation of particles in linear quantum field theories can
be understood as a mixing of positive and negative frequencies.}.  

In theories  with deformed or broken Lorentz invariance 
(e.g., due to a modified dispersion relation), 
even inertial detectors may click if they move at super-luminal velocities -- 
or, more precisely, at speeds faster than the phase velocity of the 
surrounding quantum vacuum 
fluctuations~\cite{BirrellDavies_1982, Horsley_2016, Kajuri_2016, Husain_2016, 
Stargen_2017, Louko_2018, Marino_2017, Tian_2021}. 
As a laboratory implementation, one could envision an inertial detector moving through 
a dielectric at a velocity greater than the medium speed of light. 
This phenomenon is referred to as the Ginzburg effect~\cite{Ginzburg_1986, Ginzburg_1996}
and can be regarded as a quantum analog to Cherenkov radiation
\cite{Cerenkov_1937, Afanasiev_1999, Kheirandish_2010, 
Meyer_1985, Brevik_1988, Fewster_2018, Svidzinsky_2021, Auston_1984, Stevens_2001, Leonhardt_2019} 
which is further related to quantum friction~\cite{Marino_2017, Barton_2010, Intravaia_2015, Klatt_2017}
(see also~\cite{Pendry_1997, Milton_2016, Volokitin_2006, ScheelBuhmann_2009, Maghrebi_2013,
Silveirinha_2014, Pieplow_2015, Volokitin_2016, Dedkov_2017}).

Actually observing the Ginzburg effect in a dielectric medium is hampered by the typically 
very large phase velocities.  
As a work-around idea, one could imagine using frequencies close to the medium resonance where the 
dielectric permittivity $\varepsilon(\omega)$ becomes very large and thus the phase velocity 
sufficiently small 
(see also~\cite{Svidzinsky_2019, Lannebere_2016}).
However, close to resonance, the imaginary part of $\varepsilon(\omega)$ also becomes 
very large such that dissipation 
starts to play an important role.
Therefore, studying the Ginzburg effect necessitates a consistent treatment of 
quantum fluctuations in the presence of dispersion and dissipation,
see also~\cite{Horsley_2012}.
In the following, we present such an approach based on a generalized version of the well-known 
Hopfield model.

\section{Model of 1D medium} \label{Model of 1D medium}

We describe the dispersive and dissipative medium 
by the following Lagrangian ($\hbar=c=1$)
\bea
L
&=& 
\frac{1}{2} \int dx \,
\Big\lbrace 
\left[\partial_t A(t,x)\right]^2 - \left[\partial_x A(t,x)\right]^2 
\nn
&&\qquad \hspace{1.25em}+
\left[\partial_t \Psi(t,x)\right]^2  -  \Omega^2 \Psi^2(t,x)
\nn
&&\qquad \hspace{1.25em}
+2gA(t,x)\partial_t\Psi(t,x)
\Big\rbrace
\nn
&&+ 
\frac{1}{2} \int dx \, d\xi \, \left\lbrace\left[\partial_t \Phi(t,x,\xi)\right]^2 
-  \left[\partial_\xi \Phi(t,x,\xi)\right]^2\right\rbrace
\nn
&&
+G\int dx\,\Phi(t,x,\xi=0) \partial_{t}\Psi(t,x)
\label{eq:Hopfield}
\,.
\ea
The first three lines correspond to the well-known Hopfield 
model~\cite{Hopfield_1958} in 1+1 dimensions  
--- a generalization to 3+1 dimensions will be discussed in Sec.~\ref{Model of 3D medium} 
below.
Here $A$ represents the electromagnetic vector potential
and $\Psi$ denotes the polarization of the medium
with the resonance frequency $\Omega$ while $g$ is the coupling between them. 
The associated transition dipole moments are assumed to be fixed. 

In order to include dissipation, the fourth line contains an 
environment field $\Phi$ propagating in an additional 
$\xi$ direction (which could be an internal coordinate or an external dimension, e.g., $y$). 
This field $\Phi$ can carry away energy to $\xi\to\pm\infty$ and is coupled to the medium $\Psi$ 
with the strength $G$ in the fifth line~\cite{Lang_2020}
--- providing a very simplified pathway for dissipation.
While this is not the most general approach~(see also~\cite{Huttner_1992Letter,  
Huttner_1992, Suttorp_2004, BuhmannScheel_2008, Philbin_2010}), 
it considers a simple and intuitive loss mechanism 
--- see Appendix~\ref{Comparison to other approaches} 
for a comparison with existing approaches.

The equations of motion derived from~\eqref{eq:Hopfield}
read 
\bea
\left[\partial_t^2 -  \partial_x^2 \right]  A(t,x) 
&=&  
g\partial_t\Psi(t,x)
\nn
\left[\partial_t^2 + \Omega^2\right] \Psi(t,x) &=&   
-g \partial_t  A(t,x)
-G \partial_t \Phi(t,x,0)\nn
\left[\partial_t^2 - \partial_\xi^2\right] \Phi(t,x,\xi)
&=& 
G\partial_t\Psi(t,x)\delta(\xi)
\label{eq:eom}
\,.
\ea
The last one can be solved in terms of the retarded Green's
function of the d'Alembertian in 1+1 dimensions 
\begin{equation}
\Phi(t,x,\xi) = 
\Phi_0(t,x,\xi) + \frac{G}{2}\Psi(t - \vert \xi \vert, x) \,,
\label{eq:Phi-solution}
\end{equation}
where $\Phi_0(t,x,\xi)$ denotes the general (free-field) solution of the homogeneous problem 
$[\partial_t^2 - \partial_\xi^2] \Phi_0(t,x,\xi) = 0$. 
Inserting this solution~\eqref{eq:Phi-solution} back into the second line of Eq.~\eqref{eq:eom}, 
we see that the (driven) harmonic oscillators $\Psi$ representing the medium in the usual Hopfield 
model turn into damped (and driven) harmonic oscillators, i.e.,  
$[\partial_t^2 + \Omega^2]\Psi\to[\partial_t^2 + 2\Gamma\partial_t + \Omega^2]\Psi$ with the 
effective damping rate $\Gamma=G^2/4$.
In the following, we will focus on the under-damped regime of $\Gamma < \Omega$.

Finally, incorporating the first line of Eq.~\eqref{eq:eom}, we arrive at the following 
decoupled equation for $A$ 
\bea
\left[
\left(\partial_t^2+\frac{G^2}{2}\,\partial_t+\Omega^2\right)
\left(\partial_t^2-\partial_x^2\right)
+g^2\partial_t^2 
\right]
A(t,x)
=
\nn
-gG\partial_t^2 \Phi_0(t,x,\xi=0)\,.
\quad
\label{eq:eom-A}
\ea

\section{Quantization} \label{Quantization} 

The homogeneous solution $\Phi_0$ is just a free field in 2+1 dimensions ($x$, $\xi$ and $t$), 
albeit propagating in $\xi$ direction only, and can thus be quantized in the standard manner 
\bea
\label{eq:quant-Phi0}
\hat\Phi_0(t,x,\xi)=\int\frac{dk\,d\kappa}{2\pi}
\frac{\hat b_{k\kappa}e^{i(kx+\kappa\xi-|\kappa|t)}}{\sqrt{2|\kappa|}}
+{\rm h.c.},
\ea
where the frequency $|\kappa|$ is independent of the wave number $k$ in $x$ direction
and just reflects the wave number $\kappa$ in the propagation direction $\xi$.
As usual, $\hat b_{k\kappa}^\dagger$ and $\hat b_{k\kappa}$ denote bosonic 
creation and annihilation operators. 

Inserting this expression~\eqref{eq:quant-Phi0} into the right-hand side of Eq.~\eqref{eq:eom-A} 
and omitting the homogeneous solutions of Eq.~\eqref{eq:eom-A} which are exponentially 
suppressed  with damping time determined by $G$ 
for late times (i.e., in a steady state), we find 
\bea
\hat A(t,x)
&=& 
\int\frac{d\kappa\,dk}{2\pi}\, gG \,
\frac{\kappa^2 \hat b_{k\kappa} e^{ikx-i|\kappa| t}}
{\zeta_{k\kappa}\sqrt{2|\kappa|}} 
+\rm h.c., 
\label{eq:quant-A-steady} 
\ea
where $\zeta_{k\kappa}$ represents the spectral decomposition 
\bea
\zeta_{k\kappa}=
\left[\Omega^2-\kappa^2-\frac{iG^2|\kappa|}{2}\right](k^2-\kappa^2)-g^2\kappa^2 
\,.
\label{eq:dispersion}
\ea

The integral in Eq.~\eqref{eq:quant-A-steady} is peaked around $\zeta_{k\kappa}=0$, 
which corresponds to the complex dispersion relation between the wave number $k$ 
and the effective frequency $|\kappa|$. 
In the limit $G\to 0$, this peak becomes infinitely sharp and thus the 
integrand of Eq.~\eqref{eq:quant-A-steady} is supported along solutions 
of the real dispersion relation.

The steady-state solution~\eqref{eq:quant-A-steady} for $\hat A$ 
and similar expressions for the other fields diagonalize the bi-linear 
system Hamiltonian $\hat H$ derived from the Lagrangian~\eqref{eq:Hopfield} via 
$:\!\hat H\!:\,=\int dk \, d\kappa \, |\kappa|  \hat b^{\dagger}_{k\kappa} \hat b_{k\kappa}$.

In Appendix~\ref{Moving reference frame}, we reconsider the above quantization procedure 
from the point of view of a moving inertial observer and discuss implications for the Ginzburg 
effect.

\section{Field correlations} \label{Field correlations}

From the $\hat A$ field in Eq.~\eqref{eq:quant-A-steady}, 
we arrive at the two-point function of the electric field operators $\hat E=-\partial_t\hat A$ 
in the vacuum state 
\bea
\langle0|\hat E(t,x)\hat E(t',x')|0\rangle 
=
\frac{g^2 G^2}{8 \pi^2} 
\int \frac{d\kappa\,dk}{|\zeta_{k\kappa}|^2}\, |\kappa|^5 
e^{ik\Delta x-i|\kappa|\Delta t}.
\nn
\label{eq:E-E-corr}
\ea
For small space-like distances $\Delta x=x-x'$ and $\Delta t=t-t'$,
we recover the free-field limit in absence of a medium 
$\propto[(\Delta t)^2+(\Delta x)^2]/[(\Delta t)^2-(\Delta x)^2]^2$ 
as expected. 
For large distances, on the other hand, we have
\bea
\label{eq:E-E-corr-large}
\langle0|\hat E(t,x)\hat E(t',x')|0\rangle 
=
-\frac{1}{2\pi n}
\frac{n^2(\Delta x)^2+(\Delta t)^2}{\left[n^2(\Delta x)^2 - (\Delta t)^2\right]^2} 
\nn
- 
\frac{g^2G^2}{\pi\Omega^4}\, 
\frac{|\Delta x|^3\left[n^2 (\Delta x)^2 + 5(\Delta t)^2\right]}
{\left[n^2 (\Delta x)^2 - (\Delta t)^2\right]^4}
+{\mathcal{O}}\left(|\Delta x|^{-4}\right)
\,,\,\,\qquad 
\ea
and see the impact of the medium.
The leading-order term in the first line just reflects the (low-energy) refractive index of our 
medium $n = \sqrt{1+g^2/\Omega^2}$ and is thus independent of dissipation.
Damping just affects the sub-leading corrections in the second line. 

\section{Wightman axioms} \label{Wightman axioms} 

It might be illuminating to compare the above two-point correlation~\eqref{eq:E-E-corr} to the 
Wightman functions of relativistic quantum fields and their properties, see, 
e.g.,~\cite{Wightman_1956,Strocchi_2004,WightmanStreater_1964}.
Being the Fourier transform of an essentially rational function, the expression~\eqref{eq:E-E-corr}
obviously defines a {\em tempered distribution}. 
However, unlike the correlators of relativistic quantum fields, it is not invariant under Lorentz 
boosts (in $x$ direction) because the medium defines a distinguished reference frame. 
As a related point, the Fourier transform 
of~\eqref{eq:E-E-corr} is supported in the entire 
half-space of $k_0>0$ and not just in the forward light cone $k_0\geq|k_1|=|k|$. 
Thus, the correlation function~\eqref{eq:E-E-corr} does neither satisfy {\em covariance} 
nor the {\em spectrum condition}, which will be important for the non-zero response of an 
inertial detector derived below. 

Nevertheless, the correlator~\eqref{eq:E-E-corr} is consistent with {\em locality} because 
the field commutator still vanishes at space-like separations. 
The {\em positivity} condition just reflects the Hilbert space structure of quantum theory 
and is thus also satisfied here, as can be checked easily because Eq.~\eqref{eq:E-E-corr} 
is the Fourier transform of a non-negative function. 
Another feature our steady-state correlation~\eqref{eq:E-E-corr} for the electric fields $E$
(not the vector potential $A$)  
has in common with relativistic quantum field theories is the 
{\em cluster property}, which here implies that the correlation vanishes at large space-like 
separations, cf.~Eq.~\eqref{eq:E-E-corr-large}.

\section{Response of inertial detector} \label{Response of inertial detector} 

Based on a two-point function analogous to~\eqref{eq:E-E-corr},
we may now calculate the response of a detector. 
We assume a point-like detector in form of a two-level system 
(e.g., an atom)
with the excitation energy $\omega$, 
moving along a prescribed trajectory $x[t]=vt$ with a constant velocity $v$. 
It is minimally coupled to the vector potential 
(as seen from its co-moving rest frame) with dipole coupling strength $\lambda$.
As usual, we start with a detector initially in the ground state and calculate its 
excitation probability by means of 
time-dependent perturbation theory (based on a power expansion in $\lambda$),
see, e.g.,~\cite{BirrellDavies_1982, Unruh_1984, Sriramkumar_1996, 
MartinMartinez_2018} and Appendix~\ref{Detector model}.

To this end, we parameterize the vector potential $\hat A(\gamma \tau, \gamma v \tau)$  
seen by the detector along its trajectory in terms of its proper time 
$\tau = t/\gamma$, where $\gamma$ is the Lorentz boost factor. 
Along with an additional factor $\lambda^2 \omega^2/\gamma$, the Fourier transform of the 
resulting two-point correlator 
evaluated at the frequency $-\omega$ corresponding to the detector gap then yields 
the excitation probability (to lowest order in $\lambda$). 
For relativistic quantum fields, the Wightman axioms, especially {\em covariance} and 
{\em spectrum condition}, ensure that every inertial detector displays zero response in 
the vacuum (unless it is super-luminal).

For our dielectric medium featuring dispersion and dissipation, however, these two conditions 
are violated resulting in a finite detector response for all non-zero velocities $v$. 
To lowest order in $\lambda$, we find the excitation probability per unit time
\bea
\frac{dP_\uparrow}{dt}=\frac{\lambda^2 g^2 G^2 \omega^2}{4\pi \gamma^2} 
\int\frac{d\kappa\,dk}{|\zeta_{k\kappa}|^2}\,
|\kappa|^3
\delta\left(\frac{\omega}{\gamma} + |\kappa| - k v\right) .\quad
\label{eq:detector}
\ea  
The Dirac delta just implements energy conservation in the detector frame.
Due to $\omega>0$, we only get contributions for $kv>|\kappa|$, i.e., 
when the detector velocity $|v|$ exceeds the phase velocity $|\kappa/k|$ 
of the modes inside the medium (in $x$ direction).  
As an intuitive picture, the detector overtakes the propagating quantum fluctuations such that 
it effectively perceives them as going backwards in time -- which means that the roles of the 
creation and annihilation operators are effectively reversed.  
This inversion is often referred to as the anomalous Doppler effect 
\cite{Gintsburg_1962,Ginzburg_1986,Ginzburg_1996},
which is also the underlying reason for classical (Cherenkov type) radiation phenomena.

The wave number $k$ must also satisfy $kv>\omega/\gamma$ which corresponds to rather 
large $|k|$ in the non-relativistic limit $|v|\ll1$. 
In this limit, the above expression~\eqref{eq:detector} can be power expanded in $v$ 
and the lowest order reads 
\bea
\frac{dP_\uparrow}{dt}
=
\int\limits_0^\infty\frac{d\kappa}{2\pi(\kappa+\omega)^4}\,
\frac{\lambda^2 g^2 G^2 \omega^2\kappa^3|v|^3}{(\kappa^2-\Omega^2)^2 + \kappa^2 G^4/4}
\,.
\ea  
At first sight, the factor $G^2$ in the numerator might be a bit surprising since it seems to 
suggest that the detector response vanishes in the limit $G\to0$ of zero dissipation.
However, this naive expectation is wrong: in the weak-dissipation limit $G\to0$, 
the integrand is peaked around the medium resonance $\kappa=\Omega$ and thus we find 
\bea
\label{eq:rate-detector}
\frac{dP_\uparrow}{dt}
=
\frac{\lambda^2 g^2 \omega^2}{2}\, 
\frac{\Omega}{(\Omega+\omega)^4}\,|v|^3
\,.
\ea  
As a function of $\omega$ (for fixed $\lambda$),
this response is maximized for detector frequencies $\omega$ 
close to the medium resonance $\Omega$. 
However, due to the scaling with $|v|^3$, the rate~\eqref{eq:rate-detector} rapidly decreases 
for small velocities $v$ \footnote{ 
Recall that as the damping goes to zero, the resonance gets sharper, and 
thus there always exists a regime in which the phase velocity is less than the detector velocity.
}.

\section{Model of 3D medium} \label{Model of 3D medium} 

As a step towards a realistic scenario, let us generalize our model~\eqref{eq:Hopfield}
to three spatial dimensions 
\bea
L
&=& 
\frac{1}{2} \int d^3r \,
\Big\lbrace 
\left[\partial_t \f{A}+\nabla\varphi\right]^2 - \left[\nabla\times\f{A}\right]^2 
\nn
&&\qquad+
\left[\partial_t\f{\Psi}\right]^2  -  \Omega^2\f{\Psi}^2
+2g\f{A}\cdot\partial_t\f{\Psi}
\Big\rbrace
\nn
&&+ 
\frac{1}{2} \int d^3r \, d\xi \, \left\lbrace\left[\partial_t \f{\Phi}\right]^2 
-  \left[\partial_\xi \f{\Phi}\right]^2\right\rbrace
\nn
&&
+G\int d^3r\,\f{\Phi}(t,\f{r},\xi=0)\cdot\partial_{t}\f{\Psi}(t,\f{r})
\label{eq:Hopfield-3D}
\, ,
\ea
where we now include a three-dimensional vector potential $\f{A}(t,\f{r})$ 
and corresponding scalar potential $\varphi(t,\f{r})$, as well as 
vector-valued polarization $\f{\Psi}(t,\f{r})$ and environment $\f{\Phi}(t,\f{r},\xi)$ fields.
Note that the damping field $\f{\Phi}(t,\f{r},\xi)$ still propagates in one dimension $\xi$ only.
Here $\xi$ represents an internal coordinate and constitutes an effective label for  
the pathway along which energy is carried away from the fields $\f{A}$ and $\f{\Psi}$ 
and eventually turned into heat (e.g., into phononic modes). 
The precise mechanism of energy transfer depends on the specific medium but could involve, 
for instance, many local electronic degrees of freedom which take energy from the $\f{A}$ and 
$\f{\Psi}$ fields and transfer it via Coulomb interactions.
If this re-distribution of energy occurs on length scales much shorter than the typical 
optical wavelengths of interest, picturing dissipation as energy transfer in an additional $\xi$ 
direction should be a good approximation.

Analogous to free electrodynamics, the field momenta are not fully independent 
but subject to constraints which can be treated in the Dirac 
formalism~\cite{Dirac_2001}.
The absence of $\dot\varphi$ in the above Lagrangian imposes a primary constraint 
$\Pi_\varphi=\partial{\cal L}/\partial\dot\varphi=0$.
The associated Euler-Lagrange equation yields the secondary constraint 
$\nabla\cdot\f{\Pi}_A=0$, i.e., the Gauss law. 
Since these constraints are of first class, we may establish the temporal and Coulomb 
gauge conditions $\varphi=0$ and $\nabla\cdot\f{A}=0$.
As a result, the vector potential becomes purely transversal and thus couples to the 
transversal components of $\f{\Psi}$ and $\f{\Phi}$ only. 
Within this transversal sub-space, we 
obtain basically the same
equations of motion and solutions as in one dimension. 

Thus the quantization can be accomplished in analogy to the one-dimensional case, 
but with two polarization unit vectors $\f{e}_{\fk{k},\sigma}$ obeying 
$\f{k}\cdot\f{e}_{\fk{k},\sigma}=0$ and integrations over $d^3k$ instead of $dk$. 
For example, Eq.~\eqref{eq:quant-A-steady} becomes  
\bea
\f{\hat{A}}(t,\f{r})
&=&
\sum_\sigma 
\int\frac{d\kappa\,d^3k}{(2\pi)^2}\, gG \,
\frac{\kappa^2 \hat b_{\fk{k}\kappa,\sigma} e^{i\fk{k}\cdot\fk{r}-i|\kappa| t}}
{\zeta_{\fk{k}\kappa}\sqrt{2|\kappa|}} 
\f{e}_{\fk{k},\sigma}
\nn
&&+\rm h.c.
\label{eq:quant-A-steady-3D} 
\ea
Note that the coupling $\f{A}\cdot\f{j}$ to the polarization charge current density 
$\f{j}=g\partial_t\f{\Psi}$ in Eq.~\eqref{eq:Hopfield-3D} does not take into account the 
polarization charge density $\varrho=-g\nabla\cdot\f{\Psi}$ of the medium, 
but the latter corresponds to the longitudinal sector 
and is not considered here. 

In analogy to Eq.~\eqref{eq:rate-detector} we may derive the excitation rate $\dot P_\uparrow$ 
of an inertial detector with velocity $v$ 
and obtain 
\bea
\frac{dP_\uparrow}{dt} =  \frac{\Lambda^2 g^2 \omega^2 }{4\pi}  \frac{\Omega}{(\Omega + \omega)^2} |v| 	\, , \quad
\label{eq:rate-detector3D-main}
\ea
see Appendix~\ref{Detector in 3D}. 
In contrast to the one-dimensional case~\eqref{eq:rate-detector}, 
this rate scales linearly in $|v|$, which stems from the enlarged phase-space 
volume in three dimensions.

\section{Experimental realization} \label{Experimental realization}

Let us sketch a possible experimental realization, 
see also~\cite{Ginzburg_1986, Ginzburg_1996}.
We choose crystalline silicon as a dielectric medium, which has a resonance at 
$\Omega\approx3.3~\rm eV$ where the phase velocity is reduced to $0.15~c$, 
corresponding to $\Re[n(\Omega)]\approx6.8$ and a wave number of $k\approx22.4~\rm eV$
~\cite{Palik_1985}.
In order to perceive the quantum fluctuations of this mode as a real photon, 
the detector must be faster than the phase velocity of $0.15~c$, 
so it can only be an atomic or sub-atomic particle. 

As a detector, we choose meta-stable hydrogen atoms in the 2s-state.
Thus the detector gap $\omega$ is an energy of the Balmer series starting at 
$\omega \approx 1.9~\rm eV$. 
To excite the atom, the Doppler shift $\f{v}\cdot\f{k}$ must lift a quantum fluctuation
with the negative frequency $\kappa = -\Omega$ up to the positive detector frequency $\omega$ 
(with the Lorentz factor $\gamma$)
\bea
\label{Doppler}
\f{v}\cdot\f{k}=\frac{\omega}{\gamma}+\Omega
\,,
\ea 
which is analogous to the argument of the Dirac delta function in Eq.~\eqref{eq:detector}. 
With the above values, we find that this requires velocities of $v\approx c/4$. 

Such velocities can be achieved for protons 
from an accelerator with an energy of about 
$30~\rm MeV$. 
By sending them through a foil, gas cell or jet of liquid helium, 
they may capture an electron and form 
hydrogen atoms~\cite{Macdonald_1974, Schwab_1987}. 
At such high energies, the capture cross section is quite small 
(about $10^{-29}~\rm cm^{2}$~\cite{Shakeshaft_1979, Kleber_1975})
such that only a small fraction $\ord(10^{-8})$ of the protons actually forms 
atoms. 
As a typical ion accelerator might deliver a current of about $1~\rm mA$, 
this corresponds to $\ord(10^8)$ atoms per second. 
The protons may be deflected out of the beam by a magnetic field 
such that only the hydrogen atoms remain.
After a propagation length of several meters, 
most of the hydrogen atoms will have decayed to the 1s ground state but 
some fraction (a few percent~\cite{Shakeshaft_1979}) 
will reach the 2s meta-stable state 
(with a lifetime of about $0.1 \, \rm s$). 
Atoms formed in long lived higher excited (Rydberg) 
states may be filtered out by field ionization (and the magnetic field). 

For a hydrogen atom actually moving \emph{inside} the silicon crystal, we find the 
probability of spontaneous excitation $\dot P_\uparrow$ to be of order $10^{-3}$ 
per centimeter path length. 
However, as sending an atom beam directly through the crystal is probably 
not feasible, we envision sending them through a small hole 
or along the surface instead. 
Obviously the relevant modes with $k \approx 22.4~\rm eV$ 
cannot propagate outside the crystal (i.e., in vacuum) 
but will leak out of the crystal in form of evanescent fields with an 
e-folding length of about $9~\rm nm$, which is much larger than the size of an atom,
see Appendix~\ref{Evanescent modes}.
As a result, only atoms passing the surface at distances of order $9~\rm nm$  may get excited. 
E.g., if a homogeneous beam of $\ord(10^6)$ atoms per second (in the 2s-state)
passes through a $1~\rm mm$ diameter hole in a silicon crystal, about $5\times 10^{-3}$
atoms per second and centimeter length of the medium 
should get excited to the 3p-state --- as detailed in Appendix~\ref{Order of magnitude estimates}. 

These atoms will then decay under the emission of a Lyman $\beta$-photon at $103~\rm nm$ 
after a radiative lifetime of $18~\rm ns$ which corresponds to a propagation length 
of $1.3~\rm m$ for hydrogen atoms with a velocity of $c/4$.
The detector to be placed behind the silicon crystal could thus consist of a cylindrical 
array of microchannel-plates of about $1~\rm m$ length that detect the Lyman $\beta$-photons. 

A similar experiment is possible with other atoms with resonance frequencies in the 
visible or near ultraviolet regime, however, much higher ion energies would be required 
to generate neutral atoms of velocities $v=c/4$.
For example, lithium would require about $200~\rm MeV$ Li$^{+}$ ions. 
While the relevant transition energy $\omega\approx1.8~\rm eV$ is similar to the hydrogen 
case, the final photon count might be larger (for the same initial current) 
because excitations would start from the atomic 
ground state rather than from a weakly-populated excited state.

\section{Conclusions} \label{Conclusions}

Via an extension of the well-known Hopfield model which can easily be generalized to three 
spatial dimensions, we provide a consistent microscopic treatment of quantum electrodynamics 
in a dielectric medium featuring dispersion and dissipation. 
The simple structure and well-defined asymptotic behavior (e.g., at $\xi\to\pm\infty$) 
of our model allow us to diagonalize the full system Hamiltonian in terms of steady-state solutions 
and to derive two-point functions of the quantum field operators, both 
without invoking further assumptions (such as the Markov approximation).

In contrast to most existing models for dissipative dielectrics~\cite{Huttner_1992Letter,  
Huttner_1992, Suttorp_2004, BuhmannScheel_2008, Philbin_2010}, 
our approach neither involves an empirically given dielectric permittivity $\varepsilon(\omega)$ 
nor uses infinitely many baths of harmonic oscillators as a pathway for dissipation.
Instead it provides a canonical quantization procedure for the famous Lorentz oscillator model known 
from classical electrodynamics~\cite{IbachLueth_2003, Jackson_1998} and hence constitutes 
a simple and clean approach to the quantum dynamics in dissipative media  
--- especially in case of Lorentzian resonances.

Comparison with the Wightman functions of relativistic quantum fields reveals that the 
two-point function in such a medium does neither satisfy 
\emph{covariance} nor the \emph{spectrum condition}. 
The latter property is closely related to the Ginzburg effect, i.e., the non-zero 
response of an inertial detector moving faster than the phase velocity of some mode. 
As an intuitive picture, the detector overtakes this mode and thus experiences its 
oscillation as going backwards in time -- thereby reversing the roles of positive 
and negative frequencies and thus creation and annihilation operators. 
In other words, the Doppler shift $vk$ or $\f{v}\cdot\f{k}$ lifts quantum fluctuations 
with negative frequencies up to positive frequencies which are then able to excite 
the detector. 

In complete analogy to acceleration radiation (a.k.a.~the Unruh effect),
Ginzburg type clicking of super-luminal detectors can be understood as 
the absorption of a quantum ground-state fluctuation and is always accompanied 
by the emission of a real photon into the medium.
In the frame of the detector, this emitted photon effectively carries negative energy 
(see Appendix~\ref{Moving reference frame}) which compensates the positive 
energy required to excite the detector. 
In the laboratory frame, both have positive energy and thus this process depletes the 
kinetic energy of the detector, leading to a small frictional 
force~\cite{Marino_2017, Barton_2010, Intravaia_2015, Klatt_2017}, 
which constitutes a special type of quantum friction 
(see also~\cite{Pendry_1997, Milton_2016, Volokitin_2006, ScheelBuhmann_2009, 
Maghrebi_2013, Silveirinha_2014, Pieplow_2015, Volokitin_2016, Dedkov_2017}).

Near the medium resonance, the phase velocity becomes very small, i.e., $k$ becomes 
very large, such that even low detector velocities $v$ may trigger excitations. 
For these modes, dissipation effects become important, which we take into account 
consistently with our model. 
We find that dissipation does not impede the Ginzburg effect.  

Actually, our idealized model incorporates modes with unlimited $k$ near the resonance, 
such that the Ginzburg effect would formally extend to arbitrarily small detector
velocities $v$.
However, for decreasing velocities $v$, the relevant phase space of modes with 
$\f{k}\cdot\f{v} > \omega/\gamma + \Omega$ shrinks,
which explains the scaling of the excitation probability with $|v|$ or $|v|^3$ in 
three and one dimensions, respectively. 

Of course, there are several limitations to describing real media by our simplified model.
Apart from the points already discussed above, our Lagrangian density is local in space and time 
and thus does not incorporate any wave number or frequency cut-off. 
However, the underlying approximations (e.g., the dipole approximation) limit the validity of 
this effective continuum description to finite band widths, in both wave number and frequency. 
Typically real media feature not just a single resonance frequency $\Omega$ but 
may have several resonances (e.g., described by the Sellmeier coefficients \cite{Sellmeier_1872}). 
Depending on the local environment of the atoms/molecules, their individual  
resonance frequencies $\Omega$ may also vary around a mean value, 
causing inhomogeneous broadening. 
In media where the resonance frequencies $\Omega$ correspond to gaps in the electronic band 
structure (e.g., between the valence and conduction bands in semiconductors), 
dispersion within those bands can also lead to resonance broadening and peak deformation.

Altogether, our simplified model should only be applied in a finite range of frequencies and 
up to a certain cut-off wave number $k_{\rm max}$.
Via Eq.~\eqref{Doppler}, this would then imply a minimum detector velocity $v_{\rm min}$
necessary to observe the Ginzburg effect. 
For crystalline silicon, the  wave number $k \approx 22.4 ~\rm eV$ corresponding to the 
measured refractive index $\Re[n(\Omega)]\approx 6.8$ at resonance $\Omega\approx3.3~\rm eV$
may serve as a conservative estimate for the cut-off $k_{\rm max}$.
Thus, an experimental test of the Ginzburg effect requires velocities $v \approx c/4$.
Using, for instance, fast hydrogen atoms as detectors we find that such an experiment 
is challenging but not out of reach.  

\acknowledgments 

This work was funded by the Deutsche Forschungsgemeinschaft (DFG, German Research Foundation) 
-- Project-ID 278162697 -- SFB 1242.
W.G.U.~acknowledges support from 
the  Natural Science and Engineering Research Council of Canada, 
the Canadian Institute for Advanced Research (CIfAR), 
the Hagler Institute for Advanced Research at Texas A\&M University, 
the Helmholtz Association and the Humboldt Foundation.

\appendix 

\section{Comparison to other approaches} \label{Comparison to other approaches} 

In using a scalar field $\Phi$ as an environment, our model substantially differs from existing 
works building upon the established Huttner-Barnett approach for dissipative 
media~\cite{Huttner_1992Letter, Huttner_1992}.
As opposed to our Lagrangian~\eqref{eq:Hopfield}, the Huttner-Barnett model incorporates dissipation 
by attaching infinitely many baths of harmonic oscillators with continuous eigen-frequencies $\omega$ 
and frequency dependent coupling strengths $G(\omega)$ to all points of a given medium. 
By diagonalizing the resulting Hamiltonian, 
the Huttner-Barnett model yields an effective dielectric permittivity $\varepsilon(\omega)$, 
which is connected to the coupling strengths $G(\omega)$ via some integral relation~\cite{Huttner_1992Letter}. 
Simplified versions of the Huttner-Barnett approach even yield exactly solvable Heisenberg equations of 
motion~\cite{Rosa_2010, *Rosa_2010_arxiv} 
but they typically assume special profiles for the coupling strengths $G(\omega)$. 
The successful microscopic derivation by Huttner and Barnett has further 
inspired the framework of \emph{macroscopic quantum electrodynamics}~\cite{BuhmannScheel_2008}, 
which provides a  phenomenological quantization scheme based on an 
\emph{empirically given} complex dielectric permittivity $\varepsilon(\omega)$.
In contrast to these approaches, our model just involves the three parameters 
$\Omega$, $g$ and $G$ and -- without further assumptions --  
allows for deriving an effective permittivity $\varepsilon(\omega)$ 
with a Lorentzian resonance 
(see also our previous work~\cite{Lang_2020}).
Further note that our Lagrangian~\eqref{eq:Hopfield} generates 
consistent quantum dynamics even in media with 
explicitly space-time dependent parameters $\Omega$, $g$ and $G$.

On a microscopic level, the Huttner-Barnett formalism and our approach can be understood as 
opposite idealizations of the same (realistic) loss mechanism.
As an exemplary pathway for dissipation, we consider a scenario, where the 
medium polarization $\Psi(t,x)$ --- being a smoothed average over a small 
neighborhood $\mathcal{M}_x$ of the point $x$ 
--- locally couples to many microscopic degrees of freedom 
(e.g. to the nuclei and electrons in the region $\mathcal{M}_x$). 
Via this interaction, energy can be transferred from the $A$ and $\Psi$ fields 
to a microscopic environment. 
In the Huttner-Barnett model this \emph{environment} is  
described by the aforementioned baths of harmonic oscillators 
(which all couple to the medium polarization but not to each other). 
This implicitly assumes the environmental degrees of freedom to be decoupled 
and allows them to store but not to permanently absorb energy. 
In fact, energy \emph{dissipated} to the harmonic baths can couple back into the 
$A$ and $\Psi$ fields (see also Ref.~\cite{Rosa_2010, *Rosa_2010_arxiv}).
In realistic media, however, interactions on the microscopic level
(e.g., Coulomb forces among the discrete electrons and nuclei) generally allow 
energy to be directly re-distributed between the environmental degrees of freedom. 
Weakly interacting environments may give rise to rather complicated 
dynamics --- for instance, energy transferred to the environment at one point $x$ 
might couple back into the $A$ and $\Psi$ fields elsewhere. 
We explore the opposite limit, where energy is re-distributed 
on short time-scales, such that all energy transferred to the environment is almost 
immediately lost for the fields $A$ and $\Psi$.
Under this assumption, it seems well-justified to model dissipation at 
each point $x$ with a scalar field $\Phi(t,x,\xi)$ that propagates 
in a perpendicular (and potentially artificial) $\xi$ direction.

\section{Moving reference frame} \label{Moving reference frame} 

To gain further insights into the Ginzburg mechanism, we briefly reconsider 
the quantization procedure from Sec.~\ref{Quantization} in the Lorentz boosted rest frame
of an inertial observer moving at a constant velocity $v$ along the medium. 
In the corresponding Lorentz transformation, we interpret the $A$ field as the single component 
of a vector potential oriented perpendicularly to the medium.
The polarization field $\Psi$ (up to a scalar pre-factor) corresponds to a length measured parallel to 
the $A$ field and $\Phi$ is treated either as the single component of a second vector potential parallel to $A$  
(if $\xi$ refers to a real spatial dimension) or as a scalar field (if $\xi$ constitutes an internal coordinate).

As before, our quantization procedure yields straightforward steady-state solutions 
which now produce a diagonalized Hamiltonian 
\begin{equation}
:\!\hat H\!:\,=\int dk \, d\kappa \, \gamma(|\kappa|-kv) \,  \hat b^{\dagger}_{k\kappa} \hat b_{k\kappa} 
\end{equation}
with the same creation and annihilation operators $b^{\dagger}_{k\kappa}$ and $\hat b_{k\kappa}$ 
as before but with the Lorentz boosted energy $\gamma(|\kappa|-kv)$, 
where $\gamma$ denotes the usual Lorentz factor. 
For pairs of $\kappa$ and $k$ with $|\kappa| < k v$, this boosted energy grows negative 
such that creating the excited state $b^{\dagger}_{k\kappa} |0\rangle$ 
from the quantum vacuum $|0\rangle$ could actually set free an energy $-\gamma(|\kappa|-kv)>0$
in the rest frame of the moving observer.
Absorbing this energy could make an inertial particle detector click 
--- which explains how the Ginzburg effect complies with energy conservation.

The condition $|\kappa| < k v$ is satisfied 
whenever our observer moves super-luminally with respect to the phase velocity $|\kappa/k|$ 
of the corresponding excitation $b^{\dagger}_{k\kappa}|0\rangle$.
Since relativistic quantum fields do not allow for super-luminal motion, 
the Ginzburg mechanism has no counterpart in relativistic field theories.

\section{Detector model} \label{Detector model} 

The following paragraph provides further details on the model used 
to describe the two-level photon detectors considered in Secs.~\ref{Response of inertial detector}
and~\ref{Model of 3D medium}. 
It applies the established Unruh-DeWitt model%
~\cite{BirrellDavies_1982, Unruh_1984, Sriramkumar_1996, MartinMartinez_2018} 
for point-like particle detectors in background fields 
to the concrete example of a hydrogen atom 
(which is the detector studied in Sec.~\ref{Experimental realization}).

We assume the nucleus to follow a prescribed inertial trajectory  
$x[\tau]= \gamma v \tau$ or $\vec{r}[\tau] = \gamma v \tau \vec e_z$ 
for the one-dimensional set-up in Sec.~\ref{Response of inertial detector} 
or the three-dimensional scenario of Sec.~\ref{Model of 3D medium} respectively, 
where $\tau = t/\gamma$ denotes the atom's proper time. 
In a co-moving reference frame, the electron undergoes 
dynamic motion with the position and momentum operators 
$\vec{\hat{q}}(\tau)$ and $\vec{\hat{p}}(\tau)$.  

For the one-dimensional set-up in Sec.~\ref{Response of inertial detector},
we use the following interaction Hamiltonian 
\begin{equation}
	\hat H_{\mathrm{int}}(\tau) = \frac{e_0}{m_e} \, \hat p(\tau)  \hat A(\gamma \tau, \gamma v \tau)\, ,
\end{equation}
to model minimal coupling between the atom (idealized as a point-particle) 
and the electromagnetic field along its trajectory, 
where $e_0$ denotes the elementary charge and $m_e$ corresponds to the electron mass. 
The corresponding operator for the three-dimensional case  
involves the scalar product between the 
three-dimensional momentum $\vec{\hat{p}}(\tau)$ 
and the vector potential $\f{\hat{\mathcal{A}}}(\tau)$  in the detector frame, 
which has the Cartesian components 
$\hat A_x(\gamma \tau, \gamma v \tau \vec{e}_z )$, 
$\hat A_y(\gamma \tau, \gamma v \tau \vec{e}_z )$ and 
$\gamma \hat A_z(\gamma \tau, \gamma v \tau \vec{e}_z )$ 
along the detector worldline.
Converting the interaction Hamiltonian $\hat H_{\mathrm{int}}(\tau)$ 
to the laboratory time $t = \gamma \tau$ produces an additional factor $\gamma^{-1}$~\footnote{
This factor automatically arises when re-parameterizing the Schrödinger equation 
$i \, \partial_\tau |\varphi(\tau)\rangle = \hat H(\tau) |\varphi(\tau)\rangle$ 
(here given with respect to the detector's proper time $\tau$) 
in terms of the laboratory time $t = \gamma \tau$ 
(for a more profound discussion in Heisenberg picture 
representation, see Sec. II. B. of Ref.~\cite{Brown_2013})}.
The detector itself, in principle, evolves according to the Hamiltonian 
$\hat{H}_{\mathrm{det}}$ of a hydrogen atom but we adopt a two-level 
approximation, where  $\hat{H}_{\mathrm{det}}$ just has two eigenstates 
$|0\rangle_{\mathrm{det}}$ and $|1\rangle_{\mathrm{det}}$ 
that are separated by an energy gap $\omega$. 

Following the standard treatment of Unruh-DeWitt detectors, 
we assume our two-level system and the surrounding field 
to initially occupy their respective ground states.  
To leading order time-dependent perturbation theory, 
we then calculate the probability per (laboratory) time 
$dP_\uparrow/dt$ (or $\dot P_\uparrow$) 
at which the detector gets spontaneously excited.

For the one-dimensional scenario in Sec.~\ref{Response of inertial detector}, 
the resulting expression 
\bea
\frac{dP_\uparrow}{dt} &=& \lambda^2 \omega^2  \int \frac{d\tau}{\gamma} \, 
							\langle 0 | 
										\hat A(\gamma \tau, \gamma v \tau)
										 \hat A(0, 0)
							| 0 \rangle \, 
							e^{-i \omega \tau} \quad \,						
\label{eq:detectorGeneral}
\ea  
involves a two-point function analogous to the field correlation~\eqref{eq:E-E-corr} 
and the coupling strength $\lambda$ corresponds to the magnitude of the 
transition matrix element $d^{(01)}$ of the dipole operator $e_0 \hat{q}$ 
evaluated with respect to the detector eigen-states $|0\rangle_{\mathrm{det}}$ 
and $|1\rangle_{\mathrm{det}}$.
The matrix element $q^{(01)}$ of the position operator $\hat{q}$ is related to the corresponding 
quantity for the momentum $\hat p$ via $p^{(01)} = -i m_e \omega q^{(01)}$.

In three dimensions, the dipole operator $e_0 \vec{\hat{q}}$ has matrix 
elements $d^{(01)}_i$ for all three spatial dimensions and 
$\dot P_\uparrow$ adopts the form
\bea
\frac{dP_\uparrow}{dt} &=& \omega^2 \sum_{i,j} d^{(01)}_i d^{(10)}_j \!\! \int \frac{d\tau}{\gamma} \, 
							\langle 0 | 
\hat{\mathcal{A}}_i(\tau) \hat{\mathcal{A}}_j (0) | 0 \rangle  \, 
							e^{-i \omega \tau} , \qquad \,						
\label{eq:detectorGeneral-3D}
\ea   
where $\hat{\mathcal{A}}_i(\tau)$ labels the components 
$\hat A_x(\gamma \tau, \gamma v \tau \vec{e}_z )$, 
$\hat A_y(\gamma \tau, \gamma v \tau \vec{e}_z )$ and 
$\gamma \hat A_z(\gamma \tau, \gamma v \tau \vec{e}_z )$  
of the Lorentz boosted vector potential $\f{\hat{\mathcal{A}}}(\tau)$ 
occurring before. 

After evaluating all quantum mechanical expectation values in 
Eqs.~\eqref{eq:detectorGeneral} and ~\eqref{eq:detectorGeneral-3D}, 
the $\tau$ integrals produce Dirac deltas. 
For the one-dimensional case, we arrive at the expression~\eqref{eq:detector} 
in the main text.
The more complicated expression for three dimensions  
is further analyzed in Appendix~\ref{Detector in 3D}.

\section{Detector in 3D}\label{Detector in 3D}

In order to obtain an explicit result for the excitation probability~\eqref{eq:detectorGeneral-3D}, 
we parameterize momentum space with spherical coordinates 
$k_x = k \sin{\theta} \cos{\phi}$, $k_y = k \sin{\theta} \sin{\phi}$ and $k_z = k \cos{\theta}$, 
and evaluate all quantum mechanical expectation values 
as well as the $\tau$ and $\phi$ integrals. 
After substituting $\eta = \cos{\theta}$, we obtain the expression 
\bea
\frac{dP_\uparrow}{dt}=\frac{g^2 G^2 \omega^2}{8\pi^2 \gamma^2}
	\int_{-\infty}^{\infty} d\kappa\, \int_0^\infty dk\, \int_{-1}^{1} d\eta \, \frac{|\kappa|^3 k^2}{|\zeta_{k\kappa}|^2} \nn
			\times \left[ \frac{1 + \eta^2 }{2} 
					\left(\left|d^{(01)}_x\right|^2 + \left|d^{(01)}_y\right|^2\right) + (1- \eta^2) 
							 \left|d^{(01)}_z\right|^2\right] \nn
			\times \delta\left(\frac{\omega}{\gamma} + |\kappa| - \eta k v\right) , \; \quad
\label{eq:detector-3DStart}
\ea 
which has a structure quite similar to Eq.~\eqref{eq:detector} for a one-dimensional 
medium.
The Dirac delta, once again, imposes the conservation of energy 
and the term $\eta k v = k_z v $ in its argument accounts for the projection 
$\vec{k}\cdot\vec{v}$ of the wave vector $\vec{k}$ onto the detector speed $\vec{v}$.  

Analogous to the one-dimensional case, we next evaluate the $k$ integral and 
then adopt the non-relativistic limit, 
in which the remaining $\kappa$ and $\eta$ integrals decouple. 
Using straightforward symmetry arguments, we restrict those integrals to 
the domains of positive $\kappa$ and $\eta \in [0,1]$. 
As before, the $\kappa$ integral is peaked at $\kappa = \Omega$ for weakly dissipative media  
(formally for $G \to 0$).
The $\eta$ integrals read $\int_0^1 d\eta \, \eta(1+\eta^2)/2 = 3/8$ 
for the $d^{(01)}_x$ and $d^{(01)}_y$ contributions to $\dot P_\uparrow$,  
and $\int_0^1 d\eta \, \eta(1-\eta^2)=1/4$ for the $d^{(01)}_z$ term. 
Eventually, we arrive at the following excitation probability 
for small detector speeds $v$ and sufficiently weak dissipation $G$
\bea
\frac{dP_\uparrow}{dt} =  \frac{g^2 \omega^2 }{4\pi}  \frac{\Omega}{(\Omega + \omega)^2} |v| \hspace{4cm} \nn
							\times \left[\frac{3}{8} \left(\left|d^{(01)}_x\right|^2 
														 + \left|d^{(01)}_y\right|^2\right) 
								+ \frac{1}{4} \left|d^{(01)}_z\right|^2 \right]  
						\, . \quad
\label{eq:rate-detector3D}
\ea
In SI-units (where $\Omega$, $g$ and $G^2$ would be 
measured in $\mathrm{s}^{-1}$ and $v$ in $\mathrm{m}/\mathrm{s}$), 
this expression involves another factor $\mu_0/(\hbar c^2)$ 
with $\mu_0$ denoting the magnetic vacuum permeability.
Under the assumption of identical matrix elements, $\big|d^{(01)}_i\big| = \Lambda$, 
for all spatial directions, the above result reduces to Eq.~\eqref{eq:rate-detector3D-main}
in the main text.

By substituting $\omega \mapsto -\omega$ and $v = 0$ into Eq.~\eqref{eq:detector-3DStart}, 
we can further calculate the leading (zero velocity) contribution to the 
corresponding decay rate $\dot P_\downarrow$.
In the limit of weak dissipation $G \to 0$ and for $\omega < \Omega$, we obtain
\bea
\frac{dP_\downarrow}{dt} =  \frac{\omega^3}{3\pi} \sqrt{1 + \frac{g^2}{\Omega^2 - \omega^2}} 
							\sum_{i} \left|d^{(01)}_i\right|^2 \,.
						 \quad
\label{eq:decay-detector3D}
\ea

Comparing the rates $\dot P_\uparrow$ and $\dot P_\downarrow$ 
for the specific scenario, where a hydrogen atom (with quantization axis parallel to its motion) 
is excited from the meta-stable 2s-state with quantum numbers $n = 2$, $l = m = 0$ 
to a $3p$ state with $n = 3$, $l = 1$, $m = \pm1$, yields the ratio
\bea
\label{eq:ratio-3D}
\frac{\dot P_\uparrow}{\dot P_\downarrow}
=
\frac{9}{32}\, 
\sqrt{1 - \frac{g^2}{\Omega^2-\omega^2 + g^2}}\,
\frac{g^2}{(\Omega+\omega)^2}\,\frac{\Omega}{\omega}\,|v|
\,.
\ea

Although the $d^{(01)}_x$, $d^{(01)}_y$ and $d^{(01)}_z$ contributions to 
$\dot P_\uparrow$ all seem to be of the same order of magnitude, 
additional care is required when considering realistic experiments 
because our idealized model does not yet have an ultra-violet cut-off. 
More specifically, the Dirac delta in Eq.~\eqref{eq:detector-3DStart} requires very large 
wave numbers $k = |\vec{k}|$ in the non-relativistic limit of $|v| \ll 1$, 
which do not exist in real media. 
As a simple workaround, one could truncate the $k$ integral in Eq.~\eqref{eq:detector-3DStart} 
at a suitable cut-off $k_\mathrm{max}$ 
--- corresponding, e.g., to the maximum wave number compatible with measured dispersion 
relations for the medium of interest \footnote{
This choice could still be an under-estimate because empirical data is typically discrete and 
does not necessarily capture the point of maximum $|k|$. 
}.

After introducing a wave number cut-off $k_\mathrm{max}$, the $k$ integral in Eq.~\eqref{eq:detector-3DStart},
yields an additional Heaviside function $\Theta(\eta k_{\mathrm{max}} v  - \omega/\gamma - |\kappa|)$  
coupling the remaining $\kappa$ and $\eta$ integrals. 
Contributions to the result~\eqref{eq:rate-detector3D} should hence 
stem from points $(\kappa, \eta)$ satisfying the condition 
$\eta k_{\mathrm{max}} v  > \omega/\gamma + |\kappa|$ 
(with $\gamma \approx 1$ for $|v| \ll 1$) in order to be physically reliable.
Since the $\kappa$ integrand peaks at rather large frequencies close to $\Omega$, 
the above condition will usually be satisfied only if $\eta = \cos{\theta} = k_z/k$ 
is sufficiently close to $\sign{v}$ or, more specifically, 
if $|\eta| \gtrsim \eta_{\mathrm{min}} \approx (\omega + \Omega)/(|v| k_{\mathrm{max}})$. 
This restriction decreases both pre-factors $3/8$ and $1/4$
in Eq.~\eqref{eq:rate-detector3D} but the factor $1/4$ is affected 
much more significantly because it stems from an $\eta$ integral whose integrand 
$\eta(1-\eta^2)$ vanishes at $\eta = 1$.
(The integrand $\eta(1+\eta^2)/2$  corresponding to the factor $3/8$, 
on the other hand, monotonically increases for $\eta \in [0,1]$ and adopts its maximum at $\eta = 1$.) 
In a refined version of Eq.~\eqref{eq:rate-detector3D}, the $d^{(01)}_z$ contribution 
would thus be negligible compared to the $d^{(01)}_x$ and $d^{(01)}_y$ terms 
and the pre-factor $3/8$ would be replaced with  
$3/8 - \eta_{\mathrm{min}}^2/4 - \eta_{\mathrm{min}}^4/8$
(or $0$ if $\eta_{\mathrm{min}} > 1$).

Although the result~\eqref{eq:rate-detector3D} has been derived in the non-relativistic 
limit of $|v| \ll 1$ and for vanishing dissipation $G \to 0$, it remains a reasonable approximation 
(at least for order of magnitude estimates) in scenarios with moderately relativistic detector speeds $v$ 
and weakly dissipative media. 
In case of moderately relativistic velocities (such as $v = c/4$), one could refine the above 
result by adding higher order corrections to our small-velocity expansion. 
However, compared to the lowest-order result, the first non-vanishing correction  
already involves an additional factor $v^2$ which is still small, 
such that we expect corrections on the order of percent. 
For media with finite dissipation $G$, the peak of the $\kappa$ integral 
at $|\kappa| \approx \Omega$ is not infinitely sharp 
but adopts a finite width proportional to $G^2$. 
As long as the remaining integrand does not change too much in this $\kappa$ range, 
the above result remains approximately valid.

\section{Evanescent modes}\label{Evanescent modes}

Since sending particle detectors right through bulk dielectrics  
seems unfeasible in real experiments, 
one would rather send them through a small hole 
in a crystal or along the surface of a dielectric plate. 
To understand, how such a change in geometry affects the order of 
magnitude of our results, we consider the situation 
of a dielectric half space next to a vacuum environment. 
Using the same detector trajectory $\vec{r}[t] = v t \vec{e}_z $ as 
in Appendix~\ref{Detector in 3D}, 
we will assume the dielectric to fill the region with $x < -d$
--- resulting in a constant medium-detector separation $d$. 

Inside the dielectric, the electromagnetic field is described by 
our steady-state solution~\eqref{eq:quant-A-steady-3D} 
for the vector potential $\hat{\vec{A}}$ plus
additional contributions $\hat{\vec{A}}_0$ satisfying the corresponding homogeneous 
wave equation (i.e., the three-dimensional counterpart of Eq.~\eqref{eq:eom-A} 
without the $\Phi_0$ term).  
Analogous to the one-dimensional field~\eqref{eq:quant-A-steady},
the three-dimensional steady-state solution 
$\hat{\vec{A}}$ for weakly dissipative media (with sufficiently small $G$) 
approximately satisfies the dispersion relation 
$|\kappa| = \omega_{\mathrm{med}}(|\vec{k}|)$ encoded in 
the identity $\zeta_{\vec{k} \kappa} = 0$.
The homogeneous $\hat{\vec{A}}_0$ solution exactly obeys the same identity. 
In the $y$ and $z$ directions, all contributions to the 
$\hat{\vec{A}}$ and $\hat{\vec{A}}_0$ fields oscillate with real 
wave numbers $k_y$ and $k_z$.
Outside the medium, the vector potential $\hat{\vec{A}}$ can be expanded 
in terms of exponentials $\exp\{i \tilde{\vec{k}} \cdot \vec{r} - i \omega_{\mathrm{vac}}(\tilde{\vec{k}}) t\}$  
obeying the relativistic dispersion relation 
$\omega_{\mathrm{vac}}(\tilde{\vec{k}}) = \pm \sqrt{{\tilde{\vec{k}}}^2}$.
At the medium-vacuum interface, 
translation invariance in the coordinates $t$, $y$ and $z$ 
causes modes to conserve their frequencies $\omega$ and 
wave vector components $k_y$ and $k_z$.

Analogous to the energy constraint in Eq.~\eqref{eq:detector-3DStart}, 
a detector with energy gap $\omega$ moving in vacuum close to a medium 
has to satisfy the condition 
\begin{equation}
\frac{\omega}{\gamma} + \omega_{\mathrm{vac}}(\tilde{\vec{k}}) - \tilde{k}_z v = 0 
\label{eq:energyConservationOutside}
\end{equation}
in order to get excited.
Since $\omega_{\mathrm{vac}}(\tilde{\vec{k}}) = \omega_{\mathrm{med}}(k)$, 
$k_y = \tilde k_y$ and $k_z = \tilde k_z$, this is essentially the same 
requirement as for a detector moving through a bulk medium.
For the latter scenario, we already know Ginzburg excitations 
to involve wave vectors $\vec{k}$ that are almost parallel to the detector velocity $\vec{v}$ 
(such that $k_z \approx \sign{v} |\vec{k}|$ and $k_x$, $k_y \approx 0$). 
Apart from that, $|\vec{k}|$ has to be large, 
which requires frequencies $\omega_{\mathrm{vac}}(\tilde{\vec{k}})$ 
close to the medium resonance $\Omega$. 
Therefore, the modes $\exp\{\im \tilde{\vec{k}} \cdot \vec{r} - i \omega_{\mathrm{vac}}(\tilde{\vec{k}}) t\}$  
potentially exciting our detector outside the medium need to satisfy the condition 
$\omega_{\mathrm{vac}}^2(\tilde{\vec{k}}) = \tilde k_x^2 + k_y^2 + k_z^2 \approx \Omega^2$. 
Approximating $k_y \approx 0$, we thus obtain 
$\tilde k_x \approx i \sqrt{k_z^2 - \Omega^2}$
which is purely imaginary for media with sub-luminal dispersion.  

As a result, the relevant modes seen by our detector are 
evanescent fields and undergo exponentially decay with the 
e-folding length 
\begin{equation}
\ell(k_z) = \frac{1}{\sqrt{k_z^2 - \Omega^2}} \, .
\label{eq:eFoldingLength}
\end{equation}

Previous studies of atoms moving next to medium-vacuum 
interfaces~\cite{Barton_2010, Intravaia_2015} 
have found similar scaling behavior with respect to the distance $d$ 
and further predict the Ginzburg effect  
to be exponentially suppressed for small detector speeds $v$.
By just comparing our Eqs.~\eqref{eq:energyConservationOutside} 
and~\eqref{eq:eFoldingLength}, one would draw the same conclusion here: 
If $v$ grows very small, $\tilde k_z = k_z$ needs to be large 
to facilitate Ginzburg excitations.
Thus, the e-folding length $\ell(k_z)$ shrinks  for decreasing velocities $v$. 
However, in real media, there usually exist medium-specific  
maximum values $k_{\mathrm{max}}$ for $|\vec{k}|$ and $k_z$, 
such that inertial detectors need to exceed certain  
boundary velocities $v_{\mathrm{min}}$ in order to get excited. 
Other works on quantum friction have predicted rational scaling 
$\sim v^n/d^m$ with velocity $v$ and distance $d$, 
where $n$ and $m$ denote system specific integers. 
However, these works either study other friction mechanisms, 
such as pair-wise photon production~\cite{Barton_2010, Intravaia_2015}, 
or involve \emph{detectors} with richer internal dynamics, 
such as metallic nano-particles~\cite{Pieplow_2015}.

\section{Order of magnitude estimates}\label{Order of magnitude estimates}

In Sec.~\ref{Experimental realization} of the main text, 
we propose an experimental test of the Ginzburg effect 
which uses inertial hydrogen atoms in the meta-stable 2s-state 
as photon detectors and crystalline silicon as a dielectric medium. 
The detector transition 2s $\mapsto$ 3p of minimum energy then
corresponds to a detector gap $\omega = 1.9 \, \rm eV$ and the resonance frequency of crystalline 
silicon reads $\Omega = 3.3\, \rm eV$~\cite{Palik_1985}. 
As a cut-off wave number $k_{\mathrm{max}} = 22.4 \, \rm eV$, we use the maximum value 
of $|\vec{k}|$ compatible with the real part of the 
refractive index at resonance $\Re[n(\Omega)]\approx6.8$ reported in Ref.~\cite{Palik_1985}.
The coupling strength $g$ between the electromagnetic field and the medium polarization 
can be extracted from the low-energy refractive index $n = 3.4$ of silicon 
via $n = \sqrt{1 + g^2/\Omega^2}$ 
and the parameters $d_i^{(01)}$ are just the transition dipole matrix elements 
for the $2s \mapsto 3p$ excitation in an unperturbed hydrogen atom 
(within the scope of our perturbative treatment).

For a single hydrogen atom moving through a bulk of silicon, 
Eq.~\eqref{eq:rate-detector3D} 
--- after replacing $3/8 \mapsto \max\left\{3/8 - \eta_{\mathrm{min}}^2/4 - \eta_{\mathrm{min}}^4/8,0\right\}$ 
and $1/4 \mapsto 0$ as explained in Appendix~\ref{Detector in 3D} 
--- yields an excitation probability $dP_\uparrow / d(|v| t)$ of order 
$10^{-4} \, \rm cm^{-1}$ per path $|v| t$ of the moving atom.
Although Eq.~\eqref{eq:rate-detector3D} produces a ratio $\dot P_\uparrow / |v|$ 
independent of the detector speed $v$, the velocity dependent factor 
$\eta_{\mathrm{min}} = (\omega + \Omega)/(|v| k_{\mathrm{max}})$  
enters the result when imposing a wave number cut-off $k_{\mathrm{max}}$. 
As explained in Appendix~\ref{Detector in 3D}, $\eta_{\mathrm{min}}$ becomes the lower boundary 
of the integrals $\int_0^1 d\eta \, \eta(1+\eta^2)/2$  
and $\int_0^1 d\eta \, \eta(1-\eta^2)$ and has to be less than $1$ to allow 
the Doppler shift $\vec{k}\cdot \vec{v}$ with $|\vec{k}| < k_{\mathrm{max}}$ to lift quantum 
vacuum fluctuations above the detector gap.
Here, this imposes a minimum detector velocity $v_{\mathrm{min}} \approx c/4$ that has to 
met in order to facilitate spontaneous Ginzburg excitations. 

As a proposal for a real experiment, we suggest to send a beam of atoms either parallel to a 
dielectric plate or through a hole in a silicon crystal. 
According to Appendix~\ref{Evanescent modes}, the relevant field modes 
that may excite inertial atoms in such scenarios form evanescent fields. 
For detector speeds $v$ just slightly above the velocity threshold 
$v_{\mathrm{min}} \approx c/4$,
Ginzburg excitations require the wave vector component $k_z$ in $v$ direction 
to be of order $k_{\mathrm{max}}$.
Via Eq.~\eqref{eq:eFoldingLength}, we hence obtain an e-folding length of 
$\ell({k_\mathrm{max}}) \approx 9 \, \rm nm$ (in SI units), 
which causes inertial atoms with the atom-medium separation $d$ to perceive a field amplitude 
that has been reduced by a factor of order $\exp\{- d / \ell({k_\mathrm{max}})\}$.
The corresponding excitation rate is reduced by a factor of order 
$\exp\{- 2d/\ell({k_\mathrm{max}})\}$ since $\dot P_\uparrow$ involves a 
two-point correlation of the field.

In a scenario, where a homogeneous beam of $10^6$ meta-stable H(2s)  
atoms per second passes through a hole of diameter $2R = 1 \, \rm mm$ in a silicon crystal, 
we incorporate this exponential decay by averaging the factor 
$\exp\{- 2d/\ell({k_\mathrm{max}})\}$ (where $d$ now denotes the radial separation  
from the hole's boundary) over the beam surface $\pi R^2$ 
--- producing a factor of $2\times 10^{-5}$.

Combining this factor with the flux of $10^6$ atoms per second 
and with our previous estimate for the excitation rate $dP_\uparrow / d(|v| t)$  
of a single atom in a bulk of silicon, 
we eventually estimate approximately $5\times 10^{-3}$ atoms to get spontaneously excited  
per second and centimeter path inside the hole.



\begin{thebibliography}{72}%
\makeatletter
\providecommand \@ifxundefined [1]{%
 \@ifx{#1\undefined}
}%
\providecommand \@ifnum [1]{%
 \ifnum #1\expandafter \@firstoftwo
 \else \expandafter \@secondoftwo
 \fi
}%
\providecommand \@ifx [1]{%
 \ifx #1\expandafter \@firstoftwo
 \else \expandafter \@secondoftwo
 \fi
}%
\providecommand \natexlab [1]{#1}%
\providecommand \enquote  [1]{``#1''}%
\providecommand \bibnamefont  [1]{#1}%
\providecommand \bibfnamefont [1]{#1}%
\providecommand \citenamefont [1]{#1}%
\providecommand \href@noop [0]{\@secondoftwo}%
\providecommand \href [0]{\begingroup \@sanitize@url \@href}%
\providecommand \@href[1]{\@@startlink{#1}\@@href}%
\providecommand \@@href[1]{\endgroup#1\@@endlink}%
\providecommand \@sanitize@url [0]{\catcode `\\12\catcode `\$12\catcode
  `\&12\catcode `\#12\catcode `\^12\catcode `\_12\catcode `\%12\relax}%
\providecommand \@@startlink[1]{}%
\providecommand \@@endlink[0]{}%
\providecommand \url  [0]{\begingroup\@sanitize@url \@url }%
\providecommand \@url [1]{\endgroup\@href {#1}{\urlprefix }}%
\providecommand \urlprefix  [0]{URL }%
\providecommand \Eprint [0]{\href }%
\providecommand \doibase [0]{https://doi.org/}%
\providecommand \selectlanguage [0]{\@gobble}%
\providecommand \bibinfo  [0]{\@secondoftwo}%
\providecommand \bibfield  [0]{\@secondoftwo}%
\providecommand \translation [1]{[#1]}%
\providecommand \BibitemOpen [0]{}%
\providecommand \bibitemStop [0]{}%
\providecommand \bibitemNoStop [0]{.\EOS\space}%
\providecommand \EOS [0]{\spacefactor3000\relax}%
\providecommand \BibitemShut  [1]{\csname bibitem#1\endcsname}%
\let\auto@bib@innerbib\@empty
\bibitem [{\citenamefont {Fulling}(1973)}]{Fulling_1973}%
  \BibitemOpen
  \bibfield  {author} {\bibinfo {author} {\bibfnamefont {S.~A.}\ \bibnamefont
  {Fulling}},\ }\bibfield  {title} {\bibinfo {title} {{Nonuniqueness of
  Canonical Field Quantization in Riemannian Space-Time}},\ }\href
  {https://doi.org/10.1103/PhysRevD.7.2850} {\bibfield  {journal} {\bibinfo
  {journal} {Phys. Rev. D}\ }\textbf {\bibinfo {volume} {7}},\ \bibinfo {pages}
  {2850} (\bibinfo {year} {1973})}\BibitemShut {NoStop}%
\bibitem [{\citenamefont {Hawking}(1974)}]{HawkingNature_1974}%
  \BibitemOpen
  \bibfield  {author} {\bibinfo {author} {\bibfnamefont {S.~W.}\ \bibnamefont
  {Hawking}},\ }\bibfield  {title} {\bibinfo {title} {{Black hole
  explosions?}},\ }\href {https://doi.org/10.1038/248030a0} {\bibfield
  {journal} {\bibinfo  {journal} {Nature}\ }\textbf {\bibinfo {volume} {248}},\
  \bibinfo {pages} {30} (\bibinfo {year} {1974})}\BibitemShut {NoStop}%
\bibitem [{\citenamefont {Hawking}(1975)}]{HawkingComm_1975}%
  \BibitemOpen
  \bibfield  {author} {\bibinfo {author} {\bibfnamefont {S.~W.}\ \bibnamefont
  {Hawking}},\ }\bibfield  {title} {\bibinfo {title} {{Particle creation by
  black holes}},\ }\href {https://doi.org/10.1007/BF02345020} {\bibfield
  {journal} {\bibinfo  {journal} {Comm. Math. Phys}\ }\textbf {\bibinfo
  {volume} {43}},\ \bibinfo {pages} {199} (\bibinfo {year} {1975})}\BibitemShut
  {NoStop}%
\bibitem [{\citenamefont {Schr\"odinger}(1939)}]{Schroedinger_1939}%
  \BibitemOpen
  \bibfield  {author} {\bibinfo {author} {\bibfnamefont {E.}~\bibnamefont
  {Schr\"odinger}},\ }\bibfield  {title} {\bibinfo {title} {{The proper
  vibrations of the expanding universe}},\ }\href
  {https://doi.org/10.1016/S0031-8914(39)90091-1} {\bibfield  {journal}
  {\bibinfo  {journal} {Physica}\ }\textbf {\bibinfo {volume} {6}},\ \bibinfo
  {pages} {899} (\bibinfo {year} {1939})}\BibitemShut {NoStop}%
\bibitem [{\citenamefont {Parker}(1968)}]{Parker_1968}%
  \BibitemOpen
  \bibfield  {author} {\bibinfo {author} {\bibfnamefont {L.}~\bibnamefont
  {Parker}},\ }\bibfield  {title} {\bibinfo {title} {{Particle Creation in
  Expanding Universes}},\ }\href {https://doi.org/10.1103/PhysRevLett.21.562}
  {\bibfield  {journal} {\bibinfo  {journal} {Phys. Rev. Lett.}\ }\textbf
  {\bibinfo {volume} {21}},\ \bibinfo {pages} {562} (\bibinfo {year}
  {1968})}\BibitemShut {NoStop}%
\bibitem [{\citenamefont {Parker}\ and\ \citenamefont
  {Navarro-Salas}(2017)}]{Parker_Interview}%
  \BibitemOpen
  \bibfield  {author} {\bibinfo {author} {\bibfnamefont {L.}~\bibnamefont
  {Parker}}\ and\ \bibinfo {author} {\bibfnamefont {J.}~\bibnamefont
  {Navarro-Salas}},\ }\href {https://arxiv.org/abs/1702.07132v1} {\bibinfo
  {title} {{Fifty years of cosmological particle creation}}},\ \bibinfo
  {howpublished} {arXiv:1702.07132} (\bibinfo {year} {2017})\BibitemShut
  {NoStop}%
\bibitem [{\citenamefont {Unruh}(1976)}]{Unruh_1976}%
  \BibitemOpen
  \bibfield  {author} {\bibinfo {author} {\bibfnamefont {W.~G.}\ \bibnamefont
  {Unruh}},\ }\bibfield  {title} {\bibinfo {title} {{Notes on black-hole
  evaporation}},\ }\href {https://doi.org/10.1103/PhysRevD.14.870} {\bibfield
  {journal} {\bibinfo  {journal} {Phys. Rev. D}\ }\textbf {\bibinfo {volume}
  {14}},\ \bibinfo {pages} {870} (\bibinfo {year} {1976})}\BibitemShut
  {NoStop}%
\bibitem [{\citenamefont {Unruh}\ and\ \citenamefont
  {Wald}(1984)}]{Unruh_1984}%
  \BibitemOpen
  \bibfield  {author} {\bibinfo {author} {\bibfnamefont {W.~G.}\ \bibnamefont
  {Unruh}}\ and\ \bibinfo {author} {\bibfnamefont {R.~M.}\ \bibnamefont
  {Wald}},\ }\bibfield  {title} {\bibinfo {title} {{What happens when an
  accelerating observer detects a Rindler particle}},\ }\href
  {https://doi.org/10.1103/PhysRevD.29.1047} {\bibfield  {journal} {\bibinfo
  {journal} {Phys. Rev. D}\ }\textbf {\bibinfo {volume} {29}},\ \bibinfo
  {pages} {1047} (\bibinfo {year} {1984})}\BibitemShut {NoStop}%
\bibitem [{\citenamefont {Wightman}(1956)}]{Wightman_1956}%
  \BibitemOpen
  \bibfield  {author} {\bibinfo {author} {\bibfnamefont {A.~S.}\ \bibnamefont
  {Wightman}},\ }\bibfield  {title} {\bibinfo {title} {{Quantum Field Theory in
  Terms of Vacuum Expectation Values}},\ }\href
  {https://doi.org/10.1103/PhysRev.101.860} {\bibfield  {journal} {\bibinfo
  {journal} {Phys. Rev.}\ }\textbf {\bibinfo {volume} {101}},\ \bibinfo {pages}
  {860} (\bibinfo {year} {1956})}\BibitemShut {NoStop}%
\bibitem [{\citenamefont {Strocchi}(2004)}]{Strocchi_2004}%
  \BibitemOpen
  \bibfield  {author} {\bibinfo {author} {\bibfnamefont {F.}~\bibnamefont
  {Strocchi}},\ }\bibfield  {title} {\bibinfo {title} {{Relativistic Quantum
  Mechanics and Field Theory}},\ }\href
  {https://doi.org/10.1023/B:FOOP.0000019625.30165.35} {\bibfield  {journal}
  {\bibinfo  {journal} {Found. Phys.}\ }\textbf {\bibinfo {volume} {34}},\
  \bibinfo {pages} {501} (\bibinfo {year} {2004})}\BibitemShut {NoStop}%
\bibitem [{\citenamefont {Streater}\ and\ \citenamefont
  {Wightman}(2000)}]{WightmanStreater_1964}%
  \BibitemOpen
  \bibfield  {author} {\bibinfo {author} {\bibfnamefont {R.~F.}\ \bibnamefont
  {Streater}}\ and\ \bibinfo {author} {\bibfnamefont {A.~S.}\ \bibnamefont
  {Wightman}},\ }\href@noop {} {\emph {\bibinfo {title} {{PCT, Spin and
  Statistics, and All That}}}},\ \bibinfo {edition} {{Paperback}}\ ed.\
  (\bibinfo  {publisher} {Princeton University Press},\ \bibinfo {year}
  {2000})\BibitemShut {NoStop}%
\bibitem [{Note1()}]{Note1}%
  \BibitemOpen
  \bibinfo {note} {Note that, quite generally, the creation of particles in
  linear quantum field theories can be understood as a mixing of positive and
  negative frequencies.}\BibitemShut {Stop}%
\bibitem [{\citenamefont {Birrell}\ and\ \citenamefont
  {Davies}(1982)}]{BirrellDavies_1982}%
  \BibitemOpen
  \bibfield  {author} {\bibinfo {author} {\bibfnamefont {N.~D.}\ \bibnamefont
  {Birrell}}\ and\ \bibinfo {author} {\bibfnamefont {P.~C.~W.}\ \bibnamefont
  {Davies}},\ }\href@noop {} {\emph {\bibinfo {title} {{Quantum Fields in
  Curved Space}}}},\ Cambridge Monographs on Mathematical Physics\ (\bibinfo
  {publisher} {Cambridge University Press},\ \bibinfo {year}
  {1982})\BibitemShut {NoStop}%
\bibitem [{\citenamefont {Horsley}\ and\ \citenamefont
  {Bugler-Lamb}(2016)}]{Horsley_2016}%
  \BibitemOpen
  \bibfield  {author} {\bibinfo {author} {\bibfnamefont {S.~A.~R.}\
  \bibnamefont {Horsley}}\ and\ \bibinfo {author} {\bibfnamefont
  {S.}~\bibnamefont {Bugler-Lamb}},\ }\bibfield  {title} {\bibinfo {title}
  {{Negative frequencies in wave propagation: A microscopic model}},\ }\href
  {https://doi.org/10.1103/PhysRevA.93.063828} {\bibfield  {journal} {\bibinfo
  {journal} {Phys. Rev. A}\ }\textbf {\bibinfo {volume} {93}},\ \bibinfo
  {pages} {063828} (\bibinfo {year} {2016})}\BibitemShut {NoStop}%
\bibitem [{\citenamefont {Kajuri}(2016)}]{Kajuri_2016}%
  \BibitemOpen
  \bibfield  {author} {\bibinfo {author} {\bibfnamefont {N.}~\bibnamefont
  {Kajuri}},\ }\bibfield  {title} {\bibinfo {title} {{Polymer quantization
  predicts radiation in inertial frames}},\ }\href
  {https://doi.org/10.1088/0264-9381/33/5/055007} {\bibfield  {journal}
  {\bibinfo  {journal} {Class. Quantum Gravity}\ }\textbf {\bibinfo {volume}
  {33}},\ \bibinfo {pages} {055007} (\bibinfo {year} {2016})}\BibitemShut
  {NoStop}%
\bibitem [{\citenamefont {Husain}\ and\ \citenamefont
  {Louko}(2016)}]{Husain_2016}%
  \BibitemOpen
  \bibfield  {author} {\bibinfo {author} {\bibfnamefont {V.}~\bibnamefont
  {Husain}}\ and\ \bibinfo {author} {\bibfnamefont {J.}~\bibnamefont {Louko}},\
  }\bibfield  {title} {\bibinfo {title} {{Low Energy Lorentz Violation from
  Modified Dispersion at High Energies}},\ }\href
  {https://doi.org/10.1103/PhysRevLett.116.061301} {\bibfield  {journal}
  {\bibinfo  {journal} {Phys. Rev. Lett.}\ }\textbf {\bibinfo {volume} {116}},\
  \bibinfo {pages} {061301} (\bibinfo {year} {2016})}\BibitemShut {NoStop}%
\bibitem [{\citenamefont {Stargen}\ \emph {et~al.}(2017)\citenamefont
  {Stargen}, \citenamefont {Kajuri},\ and\ \citenamefont
  {Sriramkumar}}]{Stargen_2017}%
  \BibitemOpen
  \bibfield  {author} {\bibinfo {author} {\bibfnamefont {D.~J.}\ \bibnamefont
  {Stargen}}, \bibinfo {author} {\bibfnamefont {N.}~\bibnamefont {Kajuri}},\
  and\ \bibinfo {author} {\bibfnamefont {L.}~\bibnamefont {Sriramkumar}},\
  }\bibfield  {title} {\bibinfo {title} {{Response of a rotating detector
  coupled to a polymer quantized field}},\ }\href
  {https://doi.org/10.1103/PhysRevD.96.066002} {\bibfield  {journal} {\bibinfo
  {journal} {Phys. Rev. D}\ }\textbf {\bibinfo {volume} {96}},\ \bibinfo
  {pages} {066002} (\bibinfo {year} {2017})}\BibitemShut {NoStop}%
\bibitem [{\citenamefont {Louko}\ and\ \citenamefont
  {Upton}(2018)}]{Louko_2018}%
  \BibitemOpen
  \bibfield  {author} {\bibinfo {author} {\bibfnamefont {J.}~\bibnamefont
  {Louko}}\ and\ \bibinfo {author} {\bibfnamefont {S.~D.}\ \bibnamefont
  {Upton}},\ }\bibfield  {title} {\bibinfo {title} {{Low-energy Lorentz
  violation from high-energy modified dispersion in inertial and circular
  motion}},\ }\href {https://doi.org/10.1103/PhysRevD.97.025008} {\bibfield
  {journal} {\bibinfo  {journal} {Phys. Rev. D}\ }\textbf {\bibinfo {volume}
  {97}},\ \bibinfo {pages} {025008} (\bibinfo {year} {2018})}\BibitemShut
  {NoStop}%
\bibitem [{\citenamefont {Marino}\ \emph {et~al.}(2017)\citenamefont {Marino},
  \citenamefont {Recati},\ and\ \citenamefont {Carusotto}}]{Marino_2017}%
  \BibitemOpen
  \bibfield  {author} {\bibinfo {author} {\bibfnamefont {J.}~\bibnamefont
  {Marino}}, \bibinfo {author} {\bibfnamefont {A.}~\bibnamefont {Recati}},\
  and\ \bibinfo {author} {\bibfnamefont {I.}~\bibnamefont {Carusotto}},\
  }\bibfield  {title} {\bibinfo {title} {{Casimir Forces and Quantum Friction
  from Ginzburg Radiation in Atomic Bose-Einstein Condensates}},\ }\href
  {https://doi.org/10.1103/PhysRevLett.118.045301} {\bibfield  {journal}
  {\bibinfo  {journal} {Phys. Rev. Lett.}\ }\textbf {\bibinfo {volume} {118}},\
  \bibinfo {pages} {045301} (\bibinfo {year} {2017})}\BibitemShut {NoStop}%
\bibitem [{\citenamefont {Tian}\ and\ \citenamefont {Du}(2021)}]{Tian_2021}%
  \BibitemOpen
  \bibfield  {author} {\bibinfo {author} {\bibfnamefont {Z.}~\bibnamefont
  {Tian}}\ and\ \bibinfo {author} {\bibfnamefont {J.}~\bibnamefont {Du}},\
  }\bibfield  {title} {\bibinfo {title} {{Probing low-energy Lorentz violation
  from high-energy modified dispersion in dipolar Bose-Einstein condensates}},\
  }\href {https://doi.org/10.1103/PhysRevD.103.085014} {\bibfield  {journal}
  {\bibinfo  {journal} {Phys. Rev. D}\ }\textbf {\bibinfo {volume} {103}},\
  \bibinfo {pages} {085014} (\bibinfo {year} {2021})}\BibitemShut {NoStop}%
\bibitem [{\citenamefont {Ginzburg}\ and\ \citenamefont
  {Frolov}(1986)}]{Ginzburg_1986}%
  \BibitemOpen
  \bibfield  {author} {\bibinfo {author} {\bibfnamefont {V.~L.}\ \bibnamefont
  {Ginzburg}}\ and\ \bibinfo {author} {\bibfnamefont {V.~P.}\ \bibnamefont
  {Frolov}},\ }\bibfield  {title} {\bibinfo {title} {{Excitation and emission
  of a “detector” in accelerated motion in a vacuum or in uniform motion at
  a velocity above the velocity of light in a medium}},\ }\href
  {http://jetpletters.ru/ps/1404/article_21311.shtml} {\bibfield  {journal}
  {\bibinfo  {journal} {Pis’ma Zh. Eksp. Teor. Fiz.}\ }\textbf {\bibinfo
  {volume} {43}},\ \bibinfo {pages} {265} (\bibinfo {year} {1986})},\ \bibinfo
  {note} {{[JETP Lett. {\bf 43}, 339 (1986)]}}\BibitemShut {NoStop}%
\bibitem [{\citenamefont {Ginzburg}(1996)}]{Ginzburg_1996}%
  \BibitemOpen
  \bibfield  {author} {\bibinfo {author} {\bibfnamefont {V.~L.}\ \bibnamefont
  {Ginzburg}},\ }\bibfield  {title} {\bibinfo {title} {{Radiation by uniformly
  moving sources (Vavilov-Cherenkov effect, transition radiation, and some
  other phenomena)}},\ }\href {https://doi.org/10.1070/PU1996v039n10ABEH000171}
  {\bibfield  {journal} {\bibinfo  {journal} {Phys.-Usp.}\ }\textbf {\bibinfo
  {volume} {39}},\ \bibinfo {pages} {973} (\bibinfo {year} {1996})}\BibitemShut
  {NoStop}%
\bibitem [{\citenamefont {\ifmmode~\check{C}\else
  \v{C}\fi{}erenkov}(1937)}]{Cerenkov_1937}%
  \BibitemOpen
  \bibfield  {author} {\bibinfo {author} {\bibfnamefont {P.~A.}\ \bibnamefont
  {\ifmmode~\check{C}\else \v{C}\fi{}erenkov}},\ }\bibfield  {title} {\bibinfo
  {title} {{Visible Radiation Produced by Electrons Moving in a Medium with
  Velocities Exceeding that of Light}},\ }\href
  {https://doi.org/10.1103/PhysRev.52.378} {\bibfield  {journal} {\bibinfo
  {journal} {Phys. Rev.}\ }\textbf {\bibinfo {volume} {52}},\ \bibinfo {pages}
  {378} (\bibinfo {year} {1937})}\BibitemShut {NoStop}%
\bibitem [{\citenamefont {Afanasiev}\ \emph {et~al.}(1999)\citenamefont
  {Afanasiev}, \citenamefont {Kartavenko},\ and\ \citenamefont
  {Magar}}]{Afanasiev_1999}%
  \BibitemOpen
  \bibfield  {author} {\bibinfo {author} {\bibfnamefont {G.~N.}\ \bibnamefont
  {Afanasiev}}, \bibinfo {author} {\bibfnamefont {V.~G.}\ \bibnamefont
  {Kartavenko}},\ and\ \bibinfo {author} {\bibfnamefont {E.~N.}\ \bibnamefont
  {Magar}},\ }\bibfield  {title} {\bibinfo {title} {{Vavilov–Cherenkov
  radiation in dispersive medium}},\ }\href
  {https://doi.org/https://doi.org/10.1016/S0921-4526(99)00078-2} {\bibfield
  {journal} {\bibinfo  {journal} {Physica B}\ }\textbf {\bibinfo {volume}
  {269}},\ \bibinfo {pages} {95} (\bibinfo {year} {1999})}\BibitemShut
  {NoStop}%
\bibitem [{\citenamefont {Kheirandish}\ and\ \citenamefont
  {Amooghorban}(2010)}]{Kheirandish_2010}%
  \BibitemOpen
  \bibfield  {author} {\bibinfo {author} {\bibfnamefont {F.}~\bibnamefont
  {Kheirandish}}\ and\ \bibinfo {author} {\bibfnamefont {E.}~\bibnamefont
  {Amooghorban}},\ }\bibfield  {title} {\bibinfo {title} {{Finite-temperature
  Cherenkov radiation in the presence of a magnetodielectric medium}},\ }\href
  {https://doi.org/10.1103/PhysRevA.82.042901} {\bibfield  {journal} {\bibinfo
  {journal} {Phys. Rev. A}\ }\textbf {\bibinfo {volume} {82}},\ \bibinfo
  {pages} {042901} (\bibinfo {year} {2010})}\BibitemShut {NoStop}%
\bibitem [{\citenamefont {Meyer}(1985)}]{Meyer_1985}%
  \BibitemOpen
  \bibfield  {author} {\bibinfo {author} {\bibfnamefont {P.~P.}\ \bibnamefont
  {Meyer}},\ }\bibfield  {title} {\bibinfo {title} {{A quantum Cerenkov
  effect}},\ }\href {https://doi.org/10.1088/0305-4470/18/12/021} {\bibfield
  {journal} {\bibinfo  {journal} {J. Phys. A: Math. Gen.}\ }\textbf {\bibinfo
  {volume} {18}},\ \bibinfo {pages} {2235} (\bibinfo {year}
  {1985})}\BibitemShut {NoStop}%
\bibitem [{\citenamefont {Brevik}\ and\ \citenamefont
  {Kolbenstvedt}(1988)}]{Brevik_1988}%
  \BibitemOpen
  \bibfield  {author} {\bibinfo {author} {\bibfnamefont {I.}~\bibnamefont
  {Brevik}}\ and\ \bibinfo {author} {\bibfnamefont {H.}~\bibnamefont
  {Kolbenstvedt}},\ }\bibfield  {title} {\bibinfo {title} {{Quantum point
  detector moving through a dielectric medium}},\ }\href
  {https://doi.org/https://doi.org/10.1007/BF02726563} {\bibfield  {journal}
  {\bibinfo  {journal} {Nuovo Cimento B}\ }\textbf {\bibinfo {volume} {102}},\
  \bibinfo {pages} {139} (\bibinfo {year} {1988})}\BibitemShut {NoStop}%
\bibitem [{\citenamefont {Fewster}\ \emph {et~al.}(2018)\citenamefont
  {Fewster}, \citenamefont {Pfeifer},\ and\ \citenamefont
  {Siemssen}}]{Fewster_2018}%
  \BibitemOpen
  \bibfield  {author} {\bibinfo {author} {\bibfnamefont {C.~J.}\ \bibnamefont
  {Fewster}}, \bibinfo {author} {\bibfnamefont {C.}~\bibnamefont {Pfeifer}},\
  and\ \bibinfo {author} {\bibfnamefont {D.}~\bibnamefont {Siemssen}},\
  }\bibfield  {title} {\bibinfo {title} {Quantum energy inequalities in
  premetric electrodynamics},\ }\href
  {https://doi.org/10.1103/PhysRevD.97.025019} {\bibfield  {journal} {\bibinfo
  {journal} {Phys. Rev. D}\ }\textbf {\bibinfo {volume} {97}},\ \bibinfo
  {pages} {025019} (\bibinfo {year} {2018})}\BibitemShut {NoStop}%
\bibitem [{\citenamefont {Svidzinsky}\ \emph {et~al.}(2021)\citenamefont
  {Svidzinsky}, \citenamefont {Azizi}, \citenamefont {Ben-Benjamin},
  \citenamefont {Scully},\ and\ \citenamefont {Unruh}}]{Svidzinsky_2021}%
  \BibitemOpen
  \bibfield  {author} {\bibinfo {author} {\bibfnamefont {A.}~\bibnamefont
  {Svidzinsky}}, \bibinfo {author} {\bibfnamefont {A.}~\bibnamefont {Azizi}},
  \bibinfo {author} {\bibfnamefont {J.~S.}\ \bibnamefont {Ben-Benjamin}},
  \bibinfo {author} {\bibfnamefont {M.~O.}\ \bibnamefont {Scully}},\ and\
  \bibinfo {author} {\bibfnamefont {W.}~\bibnamefont {Unruh}},\ }\bibfield
  {title} {\bibinfo {title} {{Unruh and Cherenkov Radiation from a Negative
  Frequency Perspective}},\ }\href
  {https://doi.org/10.1103/PhysRevLett.126.063603} {\bibfield  {journal}
  {\bibinfo  {journal} {Phys. Rev. Lett.}\ }\textbf {\bibinfo {volume} {126}},\
  \bibinfo {pages} {063603} (\bibinfo {year} {2021})}\BibitemShut {NoStop}%
\bibitem [{\citenamefont {Auston}\ \emph {et~al.}(1984)\citenamefont {Auston},
  \citenamefont {Cheung}, \citenamefont {Valdmanis},\ and\ \citenamefont
  {Kleinman}}]{Auston_1984}%
  \BibitemOpen
  \bibfield  {author} {\bibinfo {author} {\bibfnamefont {D.~H.}\ \bibnamefont
  {Auston}}, \bibinfo {author} {\bibfnamefont {K.~P.}\ \bibnamefont {Cheung}},
  \bibinfo {author} {\bibfnamefont {J.~A.}\ \bibnamefont {Valdmanis}},\ and\
  \bibinfo {author} {\bibfnamefont {D.~A.}\ \bibnamefont {Kleinman}},\
  }\bibfield  {title} {\bibinfo {title} {{Cherenkov Radiation from Femtosecond
  Optical Pulses in Electro-Optic Media}},\ }\href
  {https://doi.org/10.1103/PhysRevLett.53.1555} {\bibfield  {journal} {\bibinfo
   {journal} {Phys. Rev. Lett.}\ }\textbf {\bibinfo {volume} {53}},\ \bibinfo
  {pages} {1555} (\bibinfo {year} {1984})}\BibitemShut {NoStop}%
\bibitem [{\citenamefont {Stevens}\ \emph {et~al.}(2001)\citenamefont
  {Stevens}, \citenamefont {Wahlstrand}, \citenamefont {Kuhl},\ and\
  \citenamefont {Merlin}}]{Stevens_2001}%
  \BibitemOpen
  \bibfield  {author} {\bibinfo {author} {\bibfnamefont {T.~E.}\ \bibnamefont
  {Stevens}}, \bibinfo {author} {\bibfnamefont {J.~K.}\ \bibnamefont
  {Wahlstrand}}, \bibinfo {author} {\bibfnamefont {J.}~\bibnamefont {Kuhl}},\
  and\ \bibinfo {author} {\bibfnamefont {R.}~\bibnamefont {Merlin}},\
  }\bibfield  {title} {\bibinfo {title} {{Cherenkov Radiation at Speeds Below
  the Light Threshold: Phonon-Assisted Phase Matching}},\ }\href
  {https://doi.org/10.1126/science.291.5504.627} {\bibfield  {journal}
  {\bibinfo  {journal} {Science}\ }\textbf {\bibinfo {volume} {291}},\ \bibinfo
  {pages} {627} (\bibinfo {year} {2001})}\BibitemShut {NoStop}%
\bibitem [{\citenamefont {Leonhardt}\ and\ \citenamefont
  {Rosenberg}(2019)}]{Leonhardt_2019}%
  \BibitemOpen
  \bibfield  {author} {\bibinfo {author} {\bibfnamefont {U.}~\bibnamefont
  {Leonhardt}}\ and\ \bibinfo {author} {\bibfnamefont {Y.}~\bibnamefont
  {Rosenberg}},\ }\bibfield  {title} {\bibinfo {title} {Cherenkov radiation of
  light bullets},\ }\href {https://doi.org/10.1103/PhysRevA.100.063802}
  {\bibfield  {journal} {\bibinfo  {journal} {Phys. Rev. A}\ }\textbf {\bibinfo
  {volume} {100}},\ \bibinfo {pages} {063802} (\bibinfo {year}
  {2019})}\BibitemShut {NoStop}%
\bibitem [{\citenamefont {Barton}(2010)}]{Barton_2010}%
  \BibitemOpen
  \bibfield  {author} {\bibinfo {author} {\bibfnamefont {G.}~\bibnamefont
  {Barton}},\ }\bibfield  {title} {\bibinfo {title} {{On van der Waals
  friction. {II}: Between atom and half-space}},\ }\href
  {https://doi.org/10.1088/1367-2630/12/11/113045} {\bibfield  {journal}
  {\bibinfo  {journal} {New J. Phys.}\ }\textbf {\bibinfo {volume} {12}},\
  \bibinfo {pages} {113045} (\bibinfo {year} {2010})}\BibitemShut {NoStop}%
\bibitem [{\citenamefont {Intravaia}\ \emph {et~al.}(2015)\citenamefont
  {Intravaia}, \citenamefont {Mkrtchian}, \citenamefont {Buhmann},
  \citenamefont {Scheel}, \citenamefont {Dalvit},\ and\ \citenamefont
  {Henkel}}]{Intravaia_2015}%
  \BibitemOpen
  \bibfield  {author} {\bibinfo {author} {\bibfnamefont {F.}~\bibnamefont
  {Intravaia}}, \bibinfo {author} {\bibfnamefont {V.~E.}\ \bibnamefont
  {Mkrtchian}}, \bibinfo {author} {\bibfnamefont {S.~Y.}\ \bibnamefont
  {Buhmann}}, \bibinfo {author} {\bibfnamefont {S.}~\bibnamefont {Scheel}},
  \bibinfo {author} {\bibfnamefont {D.~A.~R.}\ \bibnamefont {Dalvit}},\ and\
  \bibinfo {author} {\bibfnamefont {C.}~\bibnamefont {Henkel}},\ }\bibfield
  {title} {\bibinfo {title} {Friction forces on atoms after acceleration},\
  }\href {https://doi.org/10.1088/0953-8984/27/21/214020} {\bibfield  {journal}
  {\bibinfo  {journal} {J. Phys. Condens. Matter}\ }\textbf {\bibinfo {volume}
  {27}},\ \bibinfo {pages} {214020} (\bibinfo {year} {2015})}\BibitemShut
  {NoStop}%
\bibitem [{\citenamefont {Klatt}\ \emph {et~al.}(2017)\citenamefont {Klatt},
  \citenamefont {Far\'{\i}as}, \citenamefont {Dalvit},\ and\ \citenamefont
  {Buhmann}}]{Klatt_2017}%
  \BibitemOpen
  \bibfield  {author} {\bibinfo {author} {\bibfnamefont {J.}~\bibnamefont
  {Klatt}}, \bibinfo {author} {\bibfnamefont {M.~B.}\ \bibnamefont
  {Far\'{\i}as}}, \bibinfo {author} {\bibfnamefont {D.~A.~R.}\ \bibnamefont
  {Dalvit}},\ and\ \bibinfo {author} {\bibfnamefont {S.~Y.}\ \bibnamefont
  {Buhmann}},\ }\bibfield  {title} {\bibinfo {title} {{Quantum friction in
  arbitrarily directed motion}},\ }\href
  {https://doi.org/10.1103/PhysRevA.95.052510} {\bibfield  {journal} {\bibinfo
  {journal} {Phys. Rev. A}\ }\textbf {\bibinfo {volume} {95}},\ \bibinfo
  {pages} {052510} (\bibinfo {year} {2017})}\BibitemShut {NoStop}%
\bibitem [{\citenamefont {Pendry}(1997)}]{Pendry_1997}%
  \BibitemOpen
  \bibfield  {author} {\bibinfo {author} {\bibfnamefont {J.~B.}\ \bibnamefont
  {Pendry}},\ }\bibfield  {title} {\bibinfo {title} {{Shearing the vacuum -
  quantum friction}},\ }\href {https://doi.org/10.1088/0953-8984/9/47/001}
  {\bibfield  {journal} {\bibinfo  {journal} {J. Phys. Condens. Matter}\
  }\textbf {\bibinfo {volume} {9}},\ \bibinfo {pages} {10301} (\bibinfo {year}
  {1997})}\BibitemShut {NoStop}%
\bibitem [{\citenamefont {Milton}\ \emph {et~al.}(2016)\citenamefont {Milton},
  \citenamefont {Høye},\ and\ \citenamefont {Brevik}}]{Milton_2016}%
  \BibitemOpen
  \bibfield  {author} {\bibinfo {author} {\bibfnamefont {K.~A.}\ \bibnamefont
  {Milton}}, \bibinfo {author} {\bibfnamefont {J.~S.}\ \bibnamefont {Høye}},\
  and\ \bibinfo {author} {\bibfnamefont {I.}~\bibnamefont {Brevik}},\
  }\bibfield  {title} {\bibinfo {title} {{The Reality of Casimir Friction}},\
  }\href {https://doi.org/10.3390/sym8050029} {\bibfield  {journal} {\bibinfo
  {journal} {Symmetry}\ }\textbf {\bibinfo {volume} {8}},\ \bibinfo {pages}
  {29} (\bibinfo {year} {2016})}\BibitemShut {NoStop}%
\bibitem [{\citenamefont {Volokitin}\ and\ \citenamefont
  {Persson}(2006)}]{Volokitin_2006}%
  \BibitemOpen
  \bibfield  {author} {\bibinfo {author} {\bibfnamefont {A.~I.}\ \bibnamefont
  {Volokitin}}\ and\ \bibinfo {author} {\bibfnamefont {B.~N.~J.}\ \bibnamefont
  {Persson}},\ }\bibfield  {title} {\bibinfo {title} {{Quantum field theory of
  van der Waals friction}},\ }\href
  {https://doi.org/10.1103/PhysRevB.74.205413} {\bibfield  {journal} {\bibinfo
  {journal} {Phys. Rev. B}\ }\textbf {\bibinfo {volume} {74}},\ \bibinfo
  {pages} {205413} (\bibinfo {year} {2006})}\BibitemShut {NoStop}%
\bibitem [{\citenamefont {Scheel}\ and\ \citenamefont
  {Buhmann}(2009)}]{ScheelBuhmann_2009}%
  \BibitemOpen
  \bibfield  {author} {\bibinfo {author} {\bibfnamefont {S.}~\bibnamefont
  {Scheel}}\ and\ \bibinfo {author} {\bibfnamefont {S.~Y.}\ \bibnamefont
  {Buhmann}},\ }\bibfield  {title} {\bibinfo {title} {{Casimir-Polder forces on
  moving atoms}},\ }\href {https://doi.org/10.1103/PhysRevA.80.042902}
  {\bibfield  {journal} {\bibinfo  {journal} {Phys. Rev. A}\ }\textbf {\bibinfo
  {volume} {80}},\ \bibinfo {pages} {042902} (\bibinfo {year}
  {2009})}\BibitemShut {NoStop}%
\bibitem [{\citenamefont {Maghrebi}\ \emph {et~al.}(2013)\citenamefont
  {Maghrebi}, \citenamefont {Golestanian},\ and\ \citenamefont
  {Kardar}}]{Maghrebi_2013}%
  \BibitemOpen
  \bibfield  {author} {\bibinfo {author} {\bibfnamefont {M.~F.}\ \bibnamefont
  {Maghrebi}}, \bibinfo {author} {\bibfnamefont {R.}~\bibnamefont
  {Golestanian}},\ and\ \bibinfo {author} {\bibfnamefont {M.}~\bibnamefont
  {Kardar}},\ }\bibfield  {title} {\bibinfo {title} {{Quantum Cherenkov
  radiation and noncontact friction}},\ }\href
  {https://doi.org/10.1103/PhysRevA.88.042509} {\bibfield  {journal} {\bibinfo
  {journal} {Phys. Rev. A}\ }\textbf {\bibinfo {volume} {88}},\ \bibinfo
  {pages} {042509} (\bibinfo {year} {2013})}\BibitemShut {NoStop}%
\bibitem [{\citenamefont {Silveirinha}(2014)}]{Silveirinha_2014}%
  \BibitemOpen
  \bibfield  {author} {\bibinfo {author} {\bibfnamefont {M.~G.}\ \bibnamefont
  {Silveirinha}},\ }\bibfield  {title} {\bibinfo {title} {{Theory of quantum
  friction}},\ }\href {https://doi.org/10.1088/1367-2630/16/6/063011}
  {\bibfield  {journal} {\bibinfo  {journal} {New J. Phys.}\ }\textbf {\bibinfo
  {volume} {16}},\ \bibinfo {pages} {063011} (\bibinfo {year}
  {2014})}\BibitemShut {NoStop}%
\bibitem [{\citenamefont {Pieplow}\ and\ \citenamefont
  {Henkel}(2015)}]{Pieplow_2015}%
  \BibitemOpen
  \bibfield  {author} {\bibinfo {author} {\bibfnamefont {G.}~\bibnamefont
  {Pieplow}}\ and\ \bibinfo {author} {\bibfnamefont {C.}~\bibnamefont
  {Henkel}},\ }\bibfield  {title} {\bibinfo {title} {{Cherenkov friction on a
  neutral particle moving parallel to a dielectric}},\ }\href
  {https://doi.org/10.1088/0953-8984/27/21/214001} {\bibfield  {journal}
  {\bibinfo  {journal} {J. Phys.: Condens. Matter}\ }\textbf {\bibinfo {volume}
  {27}},\ \bibinfo {pages} {214001} (\bibinfo {year} {2015})}\BibitemShut
  {NoStop}%
\bibitem [{\citenamefont {Volokitin}\ and\ \citenamefont
  {Persson}(2016)}]{Volokitin_2016}%
  \BibitemOpen
  \bibfield  {author} {\bibinfo {author} {\bibfnamefont {A.~I.}\ \bibnamefont
  {Volokitin}}\ and\ \bibinfo {author} {\bibfnamefont {B.~N.~Y.}\ \bibnamefont
  {Persson}},\ }\bibfield  {title} {\bibinfo {title} {{Quantum Cherenkov
  radiation at the motion of a small neutral particle parallel to the surface
  of a transparent dielectric}},\ }\href
  {https://doi.org/10.1134/S0021364016040147} {\bibfield  {journal} {\bibinfo
  {journal} {JETP Lett.}\ }\textbf {\bibinfo {volume} {103}},\ \bibinfo {pages}
  {228} (\bibinfo {year} {2016})}\BibitemShut {NoStop}%
\bibitem [{\citenamefont {Dedkov}\ and\ \citenamefont
  {Kyasov}(2017)}]{Dedkov_2017}%
  \BibitemOpen
  \bibfield  {author} {\bibinfo {author} {\bibfnamefont {G.~V.}\ \bibnamefont
  {Dedkov}}\ and\ \bibinfo {author} {\bibfnamefont {A.~A.}\ \bibnamefont
  {Kyasov}},\ }\bibfield  {title} {\bibinfo {title} {{Cherenkov friction and
  radiation of a neutral polarizable particle moving near a transparent
  dielectric plate}},\ }\href {https://doi.org/10.1134/S1063785017080168}
  {\bibfield  {journal} {\bibinfo  {journal} {Techn. Phys. Lett.}\ }\textbf
  {\bibinfo {volume} {43}},\ \bibinfo {pages} {760} (\bibinfo {year}
  {2017})}\BibitemShut {NoStop}%
\bibitem [{\citenamefont {Svidzinsky}(2019)}]{Svidzinsky_2019}%
  \BibitemOpen
  \bibfield  {author} {\bibinfo {author} {\bibfnamefont {A.~A.}\ \bibnamefont
  {Svidzinsky}},\ }\bibfield  {title} {\bibinfo {title} {{Excitation of a
  uniformly moving atom through vacuum fluctuations}},\ }\href
  {https://doi.org/10.1103/PhysRevResearch.1.033027} {\bibfield  {journal}
  {\bibinfo  {journal} {Phys. Rev. Research}\ }\textbf {\bibinfo {volume}
  {1}},\ \bibinfo {pages} {033027} (\bibinfo {year} {2019})}\BibitemShut
  {NoStop}%
\bibitem [{\citenamefont {Lanneb\`ere}\ and\ \citenamefont
  {Silveirinha}(2016)}]{Lannebere_2016}%
  \BibitemOpen
  \bibfield  {author} {\bibinfo {author} {\bibfnamefont {S.}~\bibnamefont
  {Lanneb\`ere}}\ and\ \bibinfo {author} {\bibfnamefont {M.~G.}\ \bibnamefont
  {Silveirinha}},\ }\bibfield  {title} {\bibinfo {title} {{Negative spontaneous
  emission by a moving two-level atom}},\ }\href
  {https://doi.org/10.1088/2040-8986/19/1/014004} {\bibfield  {journal}
  {\bibinfo  {journal} {J. Opt.}\ }\textbf {\bibinfo {volume} {19}},\ \bibinfo
  {pages} {014004} (\bibinfo {year} {2016})}\BibitemShut {NoStop}%
\bibitem [{\citenamefont {Horsley}(2012)}]{Horsley_2012}%
  \BibitemOpen
  \bibfield  {author} {\bibinfo {author} {\bibfnamefont {S.~A.~R.}\
  \bibnamefont {Horsley}},\ }\bibfield  {title} {\bibinfo {title} {{Canonical
  quantization of the electromagnetic field interacting with a moving
  dielectric medium}},\ }\href {https://doi.org/10.1103/PhysRevA.86.023830}
  {\bibfield  {journal} {\bibinfo  {journal} {Phys. Rev. A}\ }\textbf {\bibinfo
  {volume} {86}},\ \bibinfo {pages} {023830} (\bibinfo {year}
  {2012})}\BibitemShut {NoStop}%
\bibitem [{\citenamefont {Hopfield}(1958)}]{Hopfield_1958}%
  \BibitemOpen
  \bibfield  {author} {\bibinfo {author} {\bibfnamefont {J.~J.}\ \bibnamefont
  {Hopfield}},\ }\bibfield  {title} {\bibinfo {title} {{Theory of the
  Contribution of Excitons to the Complex Dielectric Constant of Crystals}},\
  }\href {https://doi.org/10.1103/PhysRev.112.1555} {\bibfield  {journal}
  {\bibinfo  {journal} {Phys. Rev.}\ }\textbf {\bibinfo {volume} {112}},\
  \bibinfo {pages} {1555} (\bibinfo {year} {1958})}\BibitemShut {NoStop}%
\bibitem [{\citenamefont {Lang}\ \emph {et~al.}(2020)\citenamefont {Lang},
  \citenamefont {Sch\"utzhold},\ and\ \citenamefont {Unruh}}]{Lang_2020}%
  \BibitemOpen
  \bibfield  {author} {\bibinfo {author} {\bibfnamefont {S.}~\bibnamefont
  {Lang}}, \bibinfo {author} {\bibfnamefont {R.}~\bibnamefont {Sch\"utzhold}},\
  and\ \bibinfo {author} {\bibfnamefont {W.~G.}\ \bibnamefont {Unruh}},\
  }\bibfield  {title} {\bibinfo {title} {Quantum radiation in dielectric media
  with dispersion and dissipation},\ }\href
  {https://doi.org/10.1103/PhysRevD.102.125020} {\bibfield  {journal} {\bibinfo
   {journal} {Phys. Rev. D}\ }\textbf {\bibinfo {volume} {102}},\ \bibinfo
  {pages} {125020} (\bibinfo {year} {2020})}\BibitemShut {NoStop}%
\bibitem [{\citenamefont {Huttner}\ and\ \citenamefont
  {Barnett}(1992{\natexlab{a}})}]{Huttner_1992Letter}%
  \BibitemOpen
  \bibfield  {author} {\bibinfo {author} {\bibfnamefont {B.}~\bibnamefont
  {Huttner}}\ and\ \bibinfo {author} {\bibfnamefont {S.~M.}\ \bibnamefont
  {Barnett}},\ }\bibfield  {title} {\bibinfo {title} {{Dispersion and Loss in a
  Hopfield Dielectric}},\ }\href {https://doi.org/10.1209/0295-5075/18/6/003}
  {\bibfield  {journal} {\bibinfo  {journal} {Europhys. Lett.}\ }\textbf
  {\bibinfo {volume} {18}},\ \bibinfo {pages} {487} (\bibinfo {year}
  {1992}{\natexlab{a}})}\BibitemShut {NoStop}%
\bibitem [{\citenamefont {Huttner}\ and\ \citenamefont
  {Barnett}(1992{\natexlab{b}})}]{Huttner_1992}%
  \BibitemOpen
  \bibfield  {author} {\bibinfo {author} {\bibfnamefont {B.}~\bibnamefont
  {Huttner}}\ and\ \bibinfo {author} {\bibfnamefont {S.~M.}\ \bibnamefont
  {Barnett}},\ }\bibfield  {title} {\bibinfo {title} {{Quantization of the
  electromagnetic field in dielectrics}},\ }\href
  {https://doi.org/10.1103/PhysRevA.46.4306} {\bibfield  {journal} {\bibinfo
  {journal} {Phys. Rev. A}\ }\textbf {\bibinfo {volume} {46}},\ \bibinfo
  {pages} {4306} (\bibinfo {year} {1992}{\natexlab{b}})}\BibitemShut {NoStop}%
\bibitem [{\citenamefont {Suttorp}\ and\ \citenamefont
  {Wubs}(2004)}]{Suttorp_2004}%
  \BibitemOpen
  \bibfield  {author} {\bibinfo {author} {\bibfnamefont {L.~G.}\ \bibnamefont
  {Suttorp}}\ and\ \bibinfo {author} {\bibfnamefont {M.}~\bibnamefont {Wubs}},\
  }\bibfield  {title} {\bibinfo {title} {{Field quantization in inhomogeneous
  absorptive dielectrics}},\ }\href
  {https://doi.org/10.1103/PhysRevA.70.013816} {\bibfield  {journal} {\bibinfo
  {journal} {Phys. Rev. A}\ }\textbf {\bibinfo {volume} {70}},\ \bibinfo
  {pages} {013816} (\bibinfo {year} {2004})}\BibitemShut {NoStop}%
\bibitem [{\citenamefont {Scheel}\ and\ \citenamefont
  {Buhmann}(2008)}]{BuhmannScheel_2008}%
  \BibitemOpen
  \bibfield  {author} {\bibinfo {author} {\bibfnamefont {S.}~\bibnamefont
  {Scheel}}\ and\ \bibinfo {author} {\bibfnamefont {S.~Y.}\ \bibnamefont
  {Buhmann}},\ }\bibfield  {title} {\bibinfo {title} {{Macroscopic Quantum
  Electrodynamics - Concepts and Applications}},\ }\href
  {http://www.physics.sk/aps/pub.php?y=2008&pub=aps-08-05} {\bibfield
  {journal} {\bibinfo  {journal} {Acta Physica Slovaca}\ }\textbf {\bibinfo
  {volume} {58}},\ \bibinfo {pages} {675} (\bibinfo {year} {2008})}\BibitemShut
  {NoStop}%
\bibitem [{\citenamefont {Philbin}(2010)}]{Philbin_2010}%
  \BibitemOpen
  \bibfield  {author} {\bibinfo {author} {\bibfnamefont {T.~G.}\ \bibnamefont
  {Philbin}},\ }\bibfield  {title} {\bibinfo {title} {{Canonical quantization
  of macroscopic electromagnetism}},\ }\href
  {https://doi.org/10.1088/1367-2630/12/12/123008} {\bibfield  {journal}
  {\bibinfo  {journal} {New J. Phys.}\ }\textbf {\bibinfo {volume} {12}},\
  \bibinfo {pages} {123008} (\bibinfo {year} {2010})}\BibitemShut {NoStop}%
\bibitem [{\citenamefont {Sriramkumar}\ and\ \citenamefont
  {Padmanabhan}(1996)}]{Sriramkumar_1996}%
  \BibitemOpen
  \bibfield  {author} {\bibinfo {author} {\bibfnamefont {L.}~\bibnamefont
  {Sriramkumar}}\ and\ \bibinfo {author} {\bibfnamefont {T.}~\bibnamefont
  {Padmanabhan}},\ }\bibfield  {title} {\bibinfo {title} {{Finite-time response
  of inertial and uniformly accelerated Unruh - DeWitt detectors}},\ }\href
  {https://doi.org/10.1088/0264-9381/13/8/005} {\bibfield  {journal} {\bibinfo
  {journal} {Class. Quantum Grav.}\ }\textbf {\bibinfo {volume} {13}},\
  \bibinfo {pages} {2061} (\bibinfo {year} {1996})}\BibitemShut {NoStop}%
\bibitem [{\citenamefont {Mart\'{\i}n-Mart\'{\i}nez}\ and\ \citenamefont
  {Rodriguez-Lopez}(2018)}]{MartinMartinez_2018}%
  \BibitemOpen
  \bibfield  {author} {\bibinfo {author} {\bibfnamefont {E.}~\bibnamefont
  {Mart\'{\i}n-Mart\'{\i}nez}}\ and\ \bibinfo {author} {\bibfnamefont
  {P.}~\bibnamefont {Rodriguez-Lopez}},\ }\bibfield  {title} {\bibinfo {title}
  {{Relativistic quantum optics: The relativistic invariance of the
  light-matter interaction models}},\ }\href
  {https://doi.org/10.1103/PhysRevD.97.105026} {\bibfield  {journal} {\bibinfo
  {journal} {Phys. Rev. D}\ }\textbf {\bibinfo {volume} {97}},\ \bibinfo
  {pages} {105026} (\bibinfo {year} {2018})}\BibitemShut {NoStop}%
\bibitem [{\citenamefont {Gintsburg}(1962)}]{Gintsburg_1962}%
  \BibitemOpen
  \bibfield  {author} {\bibinfo {author} {\bibfnamefont {M.~A.}\ \bibnamefont
  {Gintsburg}},\ }\bibfield  {title} {\bibinfo {title} {{Anomalous Doppler
  Effect in a Plasma}},\ }\href
  {http://jetp.ras.ru/cgi-bin/e/index/e/14/3/p542?a=list} {\bibfield  {journal}
  {\bibinfo  {journal} {Zh. Eksp. Teor. Fiz.}\ }\textbf {\bibinfo {volume}
  {41}},\ \bibinfo {pages} {752} (\bibinfo {year} {1962})},\ \bibinfo {note}
  {[Sov. Phys. JETP \textbf{14}, 542 (1962)]}\BibitemShut {NoStop}%
\bibitem [{Note2()}]{Note2}%
  \BibitemOpen
  \bibinfo {note} {Recall that as the damping goes to zero, the resonance gets
  sharper, and thus there always exists a regime in which the phase velocity is
  less than the detector velocity.}\BibitemShut {Stop}%
\bibitem [{\citenamefont {Dirac}(2001)}]{Dirac_2001}%
  \BibitemOpen
  \bibfield  {author} {\bibinfo {author} {\bibfnamefont {P.~A.~M.}\
  \bibnamefont {Dirac}},\ }\href@noop {} {\emph {\bibinfo {title} {Lectures on
  Quantum Mechanics}}}\ (\bibinfo  {publisher} {Dover Publications (Mineola)},\
  \bibinfo {year} {2001})\BibitemShut {NoStop}%
\bibitem [{\citenamefont {Palik}(1985)}]{Palik_1985}%
  \BibitemOpen
  \bibfield  {author} {\bibinfo {author} {\bibfnamefont {E.~D.}\ \bibnamefont
  {Palik}},\ }\href@noop {} {\emph {\bibinfo {title} {{Handbook of Optical
  Constants of Solids}}}}\ (\bibinfo  {publisher} {Academic Press, Orlando},\
  \bibinfo {year} {1985})\BibitemShut {NoStop}%
\bibitem [{\citenamefont {Macdonald}\ \emph {et~al.}(1974)\citenamefont
  {Macdonald}, \citenamefont {Cocke},\ and\ \citenamefont
  {Eidson}}]{Macdonald_1974}%
  \BibitemOpen
  \bibfield  {author} {\bibinfo {author} {\bibfnamefont {J.~R.}\ \bibnamefont
  {Macdonald}}, \bibinfo {author} {\bibfnamefont {C.~L.}\ \bibnamefont
  {Cocke}},\ and\ \bibinfo {author} {\bibfnamefont {W.~W.}\ \bibnamefont
  {Eidson}},\ }\bibfield  {title} {\bibinfo {title} {{Capture of Argon K-Shell
  Electrons by 2.5- to 12-MeV Protons}},\ }\href
  {https://doi.org/10.1103/PhysRevLett.32.648} {\bibfield  {journal} {\bibinfo
  {journal} {Phys. Rev. Lett.}\ }\textbf {\bibinfo {volume} {32}},\ \bibinfo
  {pages} {648} (\bibinfo {year} {1974})}\BibitemShut {NoStop}%
\bibitem [{\citenamefont {Schwab}\ \emph {et~al.}(1987)\citenamefont {Schwab},
  \citenamefont {Baptista}, \citenamefont {Justiniano}, \citenamefont {Schuch},
  \citenamefont {Vogt},\ and\ \citenamefont {Weber}}]{Schwab_1987}%
  \BibitemOpen
  \bibfield  {author} {\bibinfo {author} {\bibfnamefont {W.}~\bibnamefont
  {Schwab}}, \bibinfo {author} {\bibfnamefont {G.~B.}\ \bibnamefont
  {Baptista}}, \bibinfo {author} {\bibfnamefont {E.}~\bibnamefont
  {Justiniano}}, \bibinfo {author} {\bibfnamefont {R.}~\bibnamefont {Schuch}},
  \bibinfo {author} {\bibfnamefont {H.}~\bibnamefont {Vogt}},\ and\ \bibinfo
  {author} {\bibfnamefont {E.~W.}\ \bibnamefont {Weber}},\ }\bibfield  {title}
  {\bibinfo {title} {{Measurement of the total cross sections for electron
  capture of 2.0-7.5 MeV H}$^+$ {in H, H}$_2$ {and He}},\ }\href
  {https://doi.org/10.1088/0022-3700/20/12/026} {\bibfield  {journal} {\bibinfo
   {journal} {J. Phys. B}\ }\textbf {\bibinfo {volume} {20}},\ \bibinfo {pages}
  {2825} (\bibinfo {year} {1987})}\BibitemShut {NoStop}%
\bibitem [{\citenamefont {Shakeshaft}\ and\ \citenamefont
  {Spruch}(1979)}]{Shakeshaft_1979}%
  \BibitemOpen
  \bibfield  {author} {\bibinfo {author} {\bibfnamefont {R.}~\bibnamefont
  {Shakeshaft}}\ and\ \bibinfo {author} {\bibfnamefont {L.}~\bibnamefont
  {Spruch}},\ }\bibfield  {title} {\bibinfo {title} {{Mechanisms for charge
  transfer (or for the capture of any light particle) at asymptotically high
  impact velocities}},\ }\href {https://doi.org/10.1103/RevModPhys.51.369}
  {\bibfield  {journal} {\bibinfo  {journal} {Rev. Mod. Phys.}\ }\textbf
  {\bibinfo {volume} {51}},\ \bibinfo {pages} {369} (\bibinfo {year}
  {1979})}\BibitemShut {NoStop}%
\bibitem [{\citenamefont {Kleber}\ and\ \citenamefont
  {Nagarajan}(1975)}]{Kleber_1975}%
  \BibitemOpen
  \bibfield  {author} {\bibinfo {author} {\bibfnamefont {M.}~\bibnamefont
  {Kleber}}\ and\ \bibinfo {author} {\bibfnamefont {M.~A.}\ \bibnamefont
  {Nagarajan}},\ }\bibfield  {title} {\bibinfo {title} {Charge transfer in
  high-energy atomic collisions},\ }\href
  {https://doi.org/10.1088/0022-3700/8/4/024} {\bibfield  {journal} {\bibinfo
  {journal} {J. Phys. B}\ }\textbf {\bibinfo {volume} {8}},\ \bibinfo {pages}
  {643} (\bibinfo {year} {1975})}\BibitemShut {NoStop}%
\bibitem [{\citenamefont {Ibach}\ and\ \citenamefont
  {L\"uth}(2003)}]{IbachLueth_2003}%
  \BibitemOpen
  \bibfield  {author} {\bibinfo {author} {\bibfnamefont {H.}~\bibnamefont
  {Ibach}}\ and\ \bibinfo {author} {\bibfnamefont {H.}~\bibnamefont {L\"uth}},\
  }\href@noop {} {\emph {\bibinfo {title} {{Solid-State Physics. An
  Introduction to Principles of Materials Science}}}},\ \bibinfo {edition}
  {3rd}\ ed.\ (\bibinfo  {publisher} {Springer Berlin Heidelberg},\ \bibinfo
  {year} {2003})\BibitemShut {NoStop}%
\bibitem [{\citenamefont {Jackson}(1998)}]{Jackson_1998}%
  \BibitemOpen
  \bibfield  {author} {\bibinfo {author} {\bibfnamefont {J.~D.}\ \bibnamefont
  {Jackson}},\ }\href@noop {} {\emph {\bibinfo {title} {{Classical
  Electrodynamics}}}},\ \bibinfo {edition} {3rd}\ ed.\ (\bibinfo  {publisher}
  {Wiley},\ \bibinfo {year} {1998})\BibitemShut {NoStop}%
\bibitem [{\citenamefont {Sellmeier}(1872)}]{Sellmeier_1872}%
  \BibitemOpen
  \bibfield  {author} {\bibinfo {author} {\bibfnamefont {W.}~\bibnamefont
  {Sellmeier}},\ }\bibfield  {title} {\bibinfo {title} {{Ueber die durch die
  Aetherschwingungen erregten Mitschwingungen der Körpertheilchen und deren
  Rückwirkung auf die ersteren, besonders zur Erklärung der Dispersion und
  ihrer Anomalien - II. Theil}},\ }\href
  {https://gallica.bnf.fr/ark:/12148/bpt6k152316/f401.item} {\bibfield
  {journal} {\bibinfo  {journal} {Ann. Phys. (Leipzig)}\ }\textbf {\bibinfo
  {volume} {147}},\ \bibinfo {pages} {386} (\bibinfo {year}
  {1872})}\BibitemShut {NoStop}%
\bibitem [{\citenamefont {Rosa}\ \emph {et~al.}(2010)\citenamefont {Rosa},
  \citenamefont {Dalvit},\ and\ \citenamefont {Milonni}}]{Rosa_2010}%
  \BibitemOpen
  \bibfield  {author} {\bibinfo {author} {\bibfnamefont {F.~S.~S.}\
  \bibnamefont {Rosa}}, \bibinfo {author} {\bibfnamefont {D.~A.~R.}\
  \bibnamefont {Dalvit}},\ and\ \bibinfo {author} {\bibfnamefont {P.~W.}\
  \bibnamefont {Milonni}},\ }\bibinfo {title} {{Doing Physic: A Festschrift For
  Thomas Erber}}\ (\bibinfo  {publisher} {Illinois Institute of Technology ITT
  Press},\ \bibinfo {year} {2010})\ Chap.\ \bibinfo {chapter} {{Quantum Fields
  in a Dielectric: Langevin and Exact Diagonalization Approaches}}, pp.\
  \bibinfo {pages} {187--205}\BibitemShut {NoStop}%
\bibitem [{Ros(2009)}]{Rosa_2010_arxiv}%
  \BibitemOpen
  \href {https://arxiv.org/abs/0912.0279} {\bibfield  {journal} {\bibinfo
  {journal} {arXiv:0912.0279}\ } (\bibinfo {year} {2009})}\BibitemShut
  {NoStop}%
\bibitem [{Note3()}]{Note3}%
  \BibitemOpen
  \bibinfo {note} {This factor automatically arises when re-parameterizing the
  Schrödinger equation $i \protect \tmspace +\thinmuskip {.1667em} \partial
  _\tau |\varphi (\tau )\rangle = \protect \hat H(\tau ) |\varphi (\tau
  )\rangle $ (here given with respect to the detector's proper time $\tau $) in
  terms of the laboratory time $t = \gamma \tau $ (for a more profound
  discussion in Heisenberg picture representation, see Sec. II. B. of
  Ref.~\cite {Brown_2013})}\BibitemShut {NoStop}%
\bibitem [{Note4()}]{Note4}%
  \BibitemOpen
  \bibinfo {note} {This choice could still be an under-estimate because
  empirical data is typically discrete and does not necessarily capture the
  point of maximum $|k|$.}\BibitemShut {Stop}%
\bibitem [{\citenamefont {Brown}\ \emph {et~al.}(2013)\citenamefont {Brown},
  \citenamefont {Mart\'{\i}n-Mart\'{\i}nez}, \citenamefont {Menicucci},\ and\
  \citenamefont {Mann}}]{Brown_2013}%
  \BibitemOpen
  \bibfield  {author} {\bibinfo {author} {\bibfnamefont {E.~G.}\ \bibnamefont
  {Brown}}, \bibinfo {author} {\bibfnamefont {E.}~\bibnamefont
  {Mart\'{\i}n-Mart\'{\i}nez}}, \bibinfo {author} {\bibfnamefont {N.~C.}\
  \bibnamefont {Menicucci}},\ and\ \bibinfo {author} {\bibfnamefont {R.~B.}\
  \bibnamefont {Mann}},\ }\bibfield  {title} {\bibinfo {title} {{Detectors for
  probing relativistic quantum physics beyond perturbation theory}},\ }\href
  {https://doi.org/10.1103/PhysRevD.87.084062} {\bibfield  {journal} {\bibinfo
  {journal} {Phys. Rev. D}\ }\textbf {\bibinfo {volume} {87}},\ \bibinfo
  {pages} {084062} (\bibinfo {year} {2013})}\BibitemShut {NoStop}%
\end{thebibliography}
%

\end{document}